\documentclass[twocolumn]{aastex631}
\usepackage{soul}
\shorttitle{The inner flow geometry in MAXI J1535-571}
\shortauthors{Yu et al.}
\graphicspath{{./}{figures/}}
\usepackage{threeparttable}
\usepackage[section]{placeins}

\begin{document}

\title{A spectral-timing study of the inner flow geometry in MAXI J1535--571 with $Insight$-HXMT and NICER}

\correspondingauthor{Wei Yu}
\email{yuwei@ihep.ac.cn}
\correspondingauthor{Qing-cui Bu}
\email{bu@astro.uni-tuebingen.de}
\correspondingauthor{Yue Huang}
\email{huangyue@mail.ihep.ac.cn}

\author{Wei Yu}
\affiliation{Key Laboratory of Particle Astrophysics, Institute of High Energy Physics, Chinese Academy of Sciences, 19B Yuquan Road, Beijing 100049, China}
\affiliation{University of Chinese Academy of Sciences, Chinese Academy of Sciences, Beijing 100049, China}
\author{Qing-Cui Bu}
\affiliation{Institut f{\"u}r Astronomie und Astrophysik, Kepler Center for Astro and Particle Physics, Eberhard Karls Universit{\"a}t, 72076 T{\"u}bingen, Germany}

\author{He-Xin Liu}
\affiliation{Key Laboratory of Particle Astrophysics, Institute of High Energy Physics, Chinese Academy of Sciences, 19B Yuquan Road, Beijing 100049, China}
\affiliation{University of Chinese Academy of Sciences, Chinese Academy of Sciences, Beijing 100049, China}

\author{Yue Huang}
\affiliation{Key Laboratory of Particle Astrophysics, Institute of High Energy Physics, Chinese Academy of Sciences, 19B Yuquan Road, Beijing 100049, China}

\author{Liang Zhang}
\affiliation{Key Laboratory of Particle Astrophysics, Institute of High Energy Physics, Chinese Academy of Sciences, 19B Yuquan Road, Beijing 100049, China}

\author{Zi-Xu Yang}
\affiliation{Key Laboratory of Particle Astrophysics, Institute of High Energy Physics, Chinese Academy of Sciences, 19B Yuquan Road, Beijing 100049, China}
\affiliation{University of Chinese Academy of Sciences, Chinese Academy of Sciences, Beijing 100049, China}

\author{Jin-Lu Qu}
\affiliation{Key Laboratory of Particle Astrophysics, Institute of High Energy Physics, Chinese Academy of Sciences, 19B Yuquan Road, Beijing 100049, China}

\author{Shu Zhang}
\affiliation{Key Laboratory of Particle Astrophysics, Institute of High Energy Physics, Chinese Academy of Sciences, 19B Yuquan Road, Beijing 100049, China}

\author{Li-Ming Song}
\affiliation{Key Laboratory of Particle Astrophysics, Institute of High Energy Physics, Chinese Academy of Sciences, 19B Yuquan Road, Beijing 100049, China}
\affiliation{University of Chinese Academy of Sciences, Chinese Academy of Sciences, Beijing 100049, China}

\author{Shuang-Nan Zhang}
\affiliation{Key Laboratory of Particle Astrophysics, Institute of High Energy Physics, Chinese Academy of Sciences, 19B Yuquan Road, Beijing 100049, China}
\affiliation{University of Chinese Academy of Sciences, Chinese Academy of Sciences, Beijing 100049, China}

\author{Shu-Mei Jia}
\affiliation{Key Laboratory of Particle Astrophysics, Institute of High Energy Physics, Chinese Academy of Sciences, 19B Yuquan Road, Beijing 100049, China}

\author{Xiang Ma}
\affiliation{Key Laboratory of Particle Astrophysics, Institute of High Energy Physics, Chinese Academy of Sciences, 19B Yuquan Road, Beijing 100049, China}

\author{Lian Tao}
\affiliation{Key Laboratory of Particle Astrophysics, Institute of High Energy Physics, Chinese Academy of Sciences, 19B Yuquan Road, Beijing 100049, China}

\author{Ming-Yu Ge}
\affiliation{Key Laboratory of Particle Astrophysics, Institute of High Energy Physics, Chinese Academy of Sciences, 19B Yuquan Road, Beijing 100049, China}

\author{Qing-Zhong Liu}
\affiliation{Purple Mountain Observatory, Chinese Academy of Sciences, Nanjing 210008, China}

\author{Jing-Zhi Yan}
\affiliation{Purple Mountain Observatory, Chinese Academy of Sciences, Nanjing 210008, China}

\author{Xue-Lei Cao}
\affiliation{Key Laboratory of Particle Astrophysics, Institute of High Energy Physics, Chinese Academy of Sciences, 19B Yuquan Road, Beijing 100049, China}
\author{Zhi Chang}
\affiliation{Key Laboratory of Particle Astrophysics, Institute of High Energy Physics, Chinese Academy of Sciences, 19B Yuquan Road, Beijing 100049, China}
\author{Li Chen}
\affiliation{Department of Astronomy, Beijing Normal University, Beijing 100088, China}
\author{Yong Chen}
\affiliation{Key Laboratory of Particle Astrophysics, Institute of High Energy Physics, Chinese Academy of Sciences, 19B Yuquan Road, Beijing 100049, China}
\author{Yu-Peng Chen}
\affiliation{Key Laboratory of Particle Astrophysics, Institute of High Energy Physics, Chinese Academy of Sciences, 19B Yuquan Road, Beijing 100049, China}
\author{Guo-Qiang Ding}
\affiliation{Xinjiang Astronomical Observatory, Chinese Academy of Sciences, Science 1-Street, Urumqi, Xinjiang 830011, China}
\author{Ju Guan}
\affiliation{Key Laboratory of Particle Astrophysics, Institute of High Energy Physics, Chinese Academy of Sciences, 19B Yuquan Road, Beijing 100049, China}
\author{Jing Jin}
\affiliation{Key Laboratory of Particle Astrophysics, Institute of High Energy Physics, Chinese Academy of Sciences, 19B Yuquan Road, Beijing 100049, China}
\author{Ling-Da Kong}
\affiliation{Key Laboratory of Particle Astrophysics, Institute of High Energy Physics, Chinese Academy of Sciences, 19B Yuquan Road, Beijing 100049, China}
\affiliation{University of Chinese Academy of Sciences, Chinese Academy of Sciences, Beijing 100049, China}
\author{Bing Li}
\affiliation{Key Laboratory of Particle Astrophysics, Institute of High Energy Physics, Chinese Academy of Sciences, 19B Yuquan Road, Beijing 100049, China}
\author{Cheng-Kui Li}
\affiliation{Key Laboratory of Particle Astrophysics, Institute of High Energy Physics, Chinese Academy of Sciences, 19B Yuquan Road, Beijing 100049, China}
\author{Ti-Pei Li}
\affiliation{Key Laboratory of Particle Astrophysics, Institute of High Energy Physics, Chinese Academy of Sciences, 19B Yuquan Road, Beijing 100049, China}
\affiliation{Department of Astronomy, Tsinghua University, Beijing 100084, China}
\affiliation{University of Chinese Academy of Sciences, Chinese Academy of Sciences, Beijing 100049, China}
\author{Xiao-Bo Li}
\affiliation{Key Laboratory of Particle Astrophysics, Institute of High Energy Physics, Chinese Academy of Sciences, 19B Yuquan Road, Beijing 100049, China}
\author{Jin-Yuan Liao}
\affiliation{Key Laboratory of Particle Astrophysics, Institute of High Energy Physics, Chinese Academy of Sciences, 19B Yuquan Road, Beijing 100049, China}
\author{Bai-Sheng Liu}
\affiliation{Key Laboratory of Particle Astrophysics, Institute of High Energy Physics, Chinese Academy of Sciences, 19B Yuquan Road, Beijing 100049, China}
\author{Cong-Zhan Liu}
\affiliation{Key Laboratory of Particle Astrophysics, Institute of High Energy Physics, Chinese Academy of Sciences, 19B Yuquan Road, Beijing 100049, China}
\author{Fang-Jun Lu}
\affiliation{Key Laboratory of Particle Astrophysics, Institute of High Energy Physics, Chinese Academy of Sciences, 19B Yuquan Road, Beijing 100049, China}
\author{Rui-Can Ma}
\affiliation{Key Laboratory of Particle Astrophysics, Institute of High Energy Physics, Chinese Academy of Sciences, 19B Yuquan Road, Beijing 100049, China}
\affiliation{University of Chinese Academy of Sciences, Chinese Academy of Sciences, Beijing 100049, China}
\author{Jian-Yin Nie}
\affiliation{Key Laboratory of Particle Astrophysics, Institute of High Energy Physics, Chinese Academy of Sciences, 19B Yuquan Road, Beijing 100049, China}
\author{Xiao-Qin Ren}
\affiliation{Key Laboratory of Particle Astrophysics, Institute of High Energy Physics, Chinese Academy of Sciences, 19B Yuquan Road, Beijing 100049, China}
\affiliation{University of Chinese Academy of Sciences, Chinese Academy of Sciences, Beijing 100049, China}
\author{Na Sai}
\affiliation{Key Laboratory of Particle Astrophysics, Institute of High Energy Physics, Chinese Academy of Sciences, 19B Yuquan Road, Beijing 100049, China}
\affiliation{University of Chinese Academy of Sciences, Chinese Academy of Sciences, Beijing 100049, China}
\author{Ying Tan}
\affiliation{Key Laboratory of Particle Astrophysics, Institute of High Energy Physics, Chinese Academy of Sciences, 19B Yuquan Road, Beijing 100049, China}
\author{You-Li Tuo}
\affiliation{Key Laboratory of Particle Astrophysics, Institute of High Energy Physics, Chinese Academy of Sciences, 19B Yuquan Road, Beijing 100049, China}
\affiliation{University of Chinese Academy of Sciences, Chinese Academy of Sciences, Beijing 100049, China}
\author{Ling-Jun Wang}
\affiliation{Key Laboratory of Particle Astrophysics, Institute of High Energy Physics, Chinese Academy of Sciences, 19B Yuquan Road, Beijing 100049, China}
\author{Peng-Ju Wang}
\affiliation{Key Laboratory of Particle Astrophysics, Institute of High Energy Physics, Chinese Academy of Sciences, 19B Yuquan Road, Beijing 100049, China}
\affiliation{University of Chinese Academy of Sciences, Chinese Academy of Sciences, Beijing 100049, China}
\author{Bai-Yang Wu}
\affiliation{Key Laboratory of Particle Astrophysics, Institute of High Energy Physics, Chinese Academy of Sciences, 19B Yuquan Road, Beijing 100049, China}
\affiliation{University of Chinese Academy of Sciences, Chinese Academy of Sciences, Beijing 100049, China}
\author{Guang-Cheng Xiao}
\affiliation{Key Laboratory of Particle Astrophysics, Institute of High Energy Physics, Chinese Academy of Sciences, 19B Yuquan Road, Beijing 100049, China}
\affiliation{University of Chinese Academy of Sciences, Chinese Academy of Sciences, Beijing 100049, China}
\author{Qian-Qing Yin}
\affiliation{Key Laboratory of Particle Astrophysics, Institute of High Energy Physics, Chinese Academy of Sciences, 19B Yuquan Road, Beijing 100049, China}
\author{Yuan You}
\affiliation{Key Laboratory of Particle Astrophysics, Institute of High Energy Physics, Chinese Academy of Sciences, 19B Yuquan Road, Beijing 100049, China}
\affiliation{University of Chinese Academy of Sciences, Chinese Academy of Sciences, Beijing 100049, China}
\author{Juan Zhang}
\affiliation{Key Laboratory of Particle Astrophysics, Institute of High Energy Physics, Chinese Academy of Sciences, 19B Yuquan Road, Beijing 100049, China}
\author{Peng Zhang}
\affiliation{Key Laboratory of Particle Astrophysics, Institute of High Energy Physics, Chinese Academy of Sciences, 19B Yuquan Road, Beijing 100049, China}
\author{Wei Zhang}
\affiliation{Key Laboratory of Particle Astrophysics, Institute of High Energy Physics, Chinese Academy of Sciences, 19B Yuquan Road, Beijing 100049, China}
\author{Hai-Sheng Zhao}
\affiliation{Key Laboratory of Particle Astrophysics, Institute of High Energy Physics, Chinese Academy of Sciences, 19B Yuquan Road, Beijing 100049, China}
\author{Shi-Jie Zheng}
\affiliation{Key Laboratory of Particle Astrophysics, Institute of High Energy Physics, Chinese Academy of Sciences, 19B Yuquan Road, Beijing 100049, China}
\author{Deng-Ke Zhou}
\affiliation{Key Laboratory of Particle Astrophysics, Institute of High Energy Physics, Chinese Academy of Sciences, 19B Yuquan Road, Beijing 100049, China}
\affiliation{University of Chinese Academy of Sciences, Chinese Academy of Sciences, Beijing 100049, China}

%% Note that the \and command from previous versions of AASTeX is now
%% depreciated in this version as it is no longer necessary. AASTeX 
%% automatically takes care of all commas and "and"s between authors names.

%% AASTeX 6.31 has the new \collaboration and \nocollaboration commands to
%% provide the collaboration status of a group of authors. These commands 
%% can be used either before or after the list of corresponding authors. The
%% argument for \collaboration is the collaboration identifier. Authors are
%% encouraged to surround collaboration identifiers with ()s. The 
%% \nocollaboration command takes no argument and exists to indicate that
%% the nearby authors are not part of surrounding collaborations.

%% Mark off the abstract in the ``abstract'' environment. 
\begin{abstract}

We have performed a spectral-timing analysis on the black hole X-ray binary MAXI J1535--571 during its 2017 outburst, with the aim of exploring the evolution of the inner accretion flow geometry. X-ray reverberation lags are observed in the hard-intermediate state (HIMS) and soft-intermediate state (SIMS) of the outburst. During the HIMS, the characteristic frequency of the reverberation lags $\nu_0$ (the frequency at which the soft lag turns to zero in the lag-frequency spectra) increases when the spectrum softens. This reflects a reduction of the spatial distance between the corona and accretion disc, when assuming the measured time lags are associated with the light travel time. We also find a strong correlation between $\nu_0$ and type-C Quasi Periodic Oscillation (QPO) centroid frequency $\nu_{QPO}$, which can be well explained by the Lense-Thirring (L-T) precession model under a truncated disk geometry. Despite the degeneracy in the spectral modellings, our results suggest that the accretion disc is largely truncated in the low hard state (LHS), and moves inward as the spectrum softens. Combine the spectral modelling results with the $\nu_0$ - $\nu_{QPO}$ evolution, we are inclined to believe that this source probably have a truncated disk geometry in the hard state.

\end{abstract}

%% Keywords should appear after the \end{abstract} command. 
%% The AAS Journals now uses Unified Astronomy Thesaurus concepts:
%% https://astrothesaurus.org
%% You will be asked to selected these concepts during the submission process
%% but this old "keyword" functionality is maintained in case authors want
%% to include these concepts in their preprints.

%% From the front matter, we move on to the body of the paper.
%% Sections are demarcated by \section and \subsection, respectively.
%% Observe the use of the LaTeX \label
%% command after the \subsection to give a symbolic KEY to the
%% subsection for cross-referencing in a \ref command.
%% You can use LaTeX's \ref and \label commands to keep track of
%% cross-references to sections, equations, tables, and figures.
%% That way, if you change the order of any elements, LaTeX will
%% automatically renumber them.
%%
%% We recommend that authors also use the natbib \citep
%% and \citet commands to identify citations.  The citations are
%% tied to the reference list via symbolic KEYs. The KEY corresponds
%% to the KEY in the \bibitem in the reference list below. 

\section{Introduction} \label{sec:intro}

\begin{figure*}
	\includegraphics[width=\columnwidth]{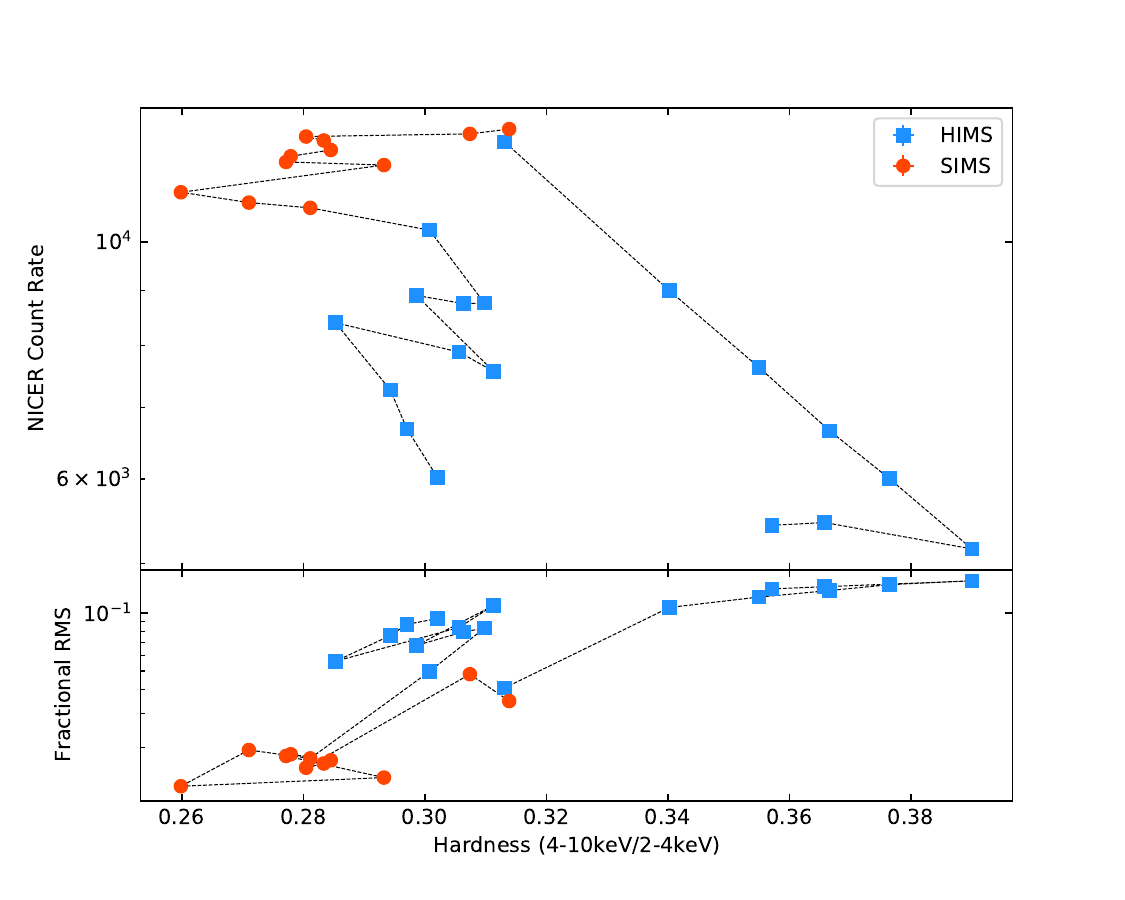}
    \includegraphics[width=\columnwidth]{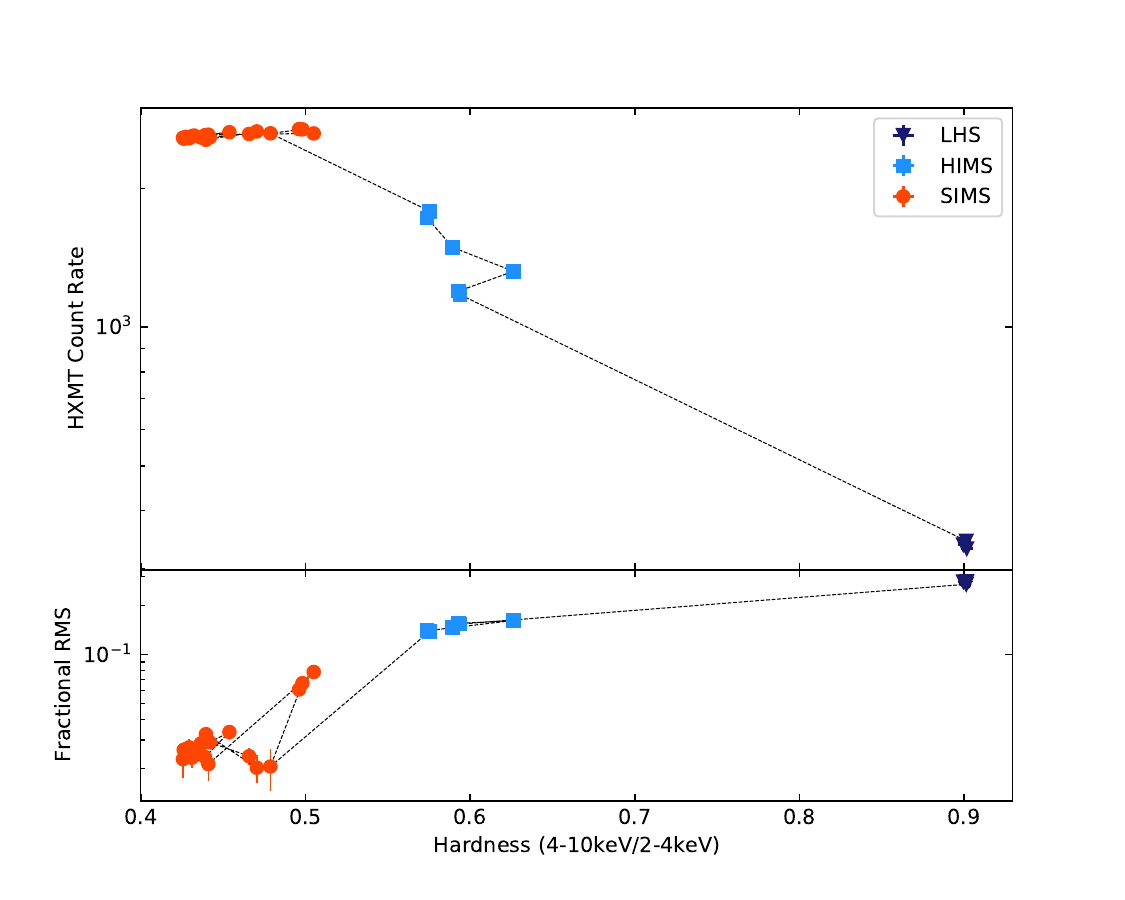}
    \caption{The hardness-intensity diagrams (HIDs) and hardness-rms diagrams (HRDs) of MAXI J1535-571. Left panel: \emph{NICER} HID\&HRD. Intensity is the count rate in 0.2-12.0 keV. Hardness is defined as the ratio of count rates between 4.0-10.0 keV and 2.0-4.0 keV. Right panel: \emph{Insight}-HXMT HID\&HRD. Intensity is the count rate in 1-10.0 keV from LE, while the hardness is defined as 4.0-10.0 keV to 2.0-4.0 keV counts ratio. Fractional averaged rms corresponds to the frequency range 0.01-64 Hz to the full energy range. Orange dots, blue squares and navy triangles represent the SIMS,  HIMS, LHS, respectively.}
    \label{hid}
\end{figure*}

Black hole low mass X-ray binaries (BH-LMXB) are mostly transient systems, in which a black hole accretes matters from its companion star via an accretion disc \citep{shakura1973black}. Black hole transients (BHTs) spend most of their lifetimes in a quiescent state, and show occasional outbursts that last from weeks to months. The outbursts could be triggered due to the instability of the system \citep{cannizzo1995accretion, lasota2001disc}. During an outburst, the source luminosity can reach the Eddington limit, while both the energy spectral properties and fast variability change dramatically, allowing for the classification of different spectral states \citep{belloni2010states}.

A unified pattern of X-ray spectral evolution of BHTs is found in most systems during an outburst, which is known as the hardness-intensity diagram (HID). The system transitions from the quiescent state to the low hard state (LHS) at the initial of the outburst. The X-ray emission during the LHS is dominated by non-thermal coronal photons, which are thought to arise from the inverse Compton scattering between the soft disk photons and the hot electrons in the corona. The X-ray spectrum in this state can be described by a phenomenological power-law with a high energy cutoff \citep{zdziarski2004radiative, remillard2006x, done2007modelling}. Strong band-limited noise and low-frequency quasi-periodic oscillations (LFQPOs) are detected in the power density spectrum (PDS). As the luminosity gradually increases, the source will evolve into the high soft state (HSS) where the spectrum is dominated by the thermal disk emission. The X-ray spectrum in the HSS can be well described by a multi-temperature disk-blackbody component \citep{remillard2006x, you2016testing}, while a power-law shaped red noise is observed in the corresponding PDS. The transitions between the LHS and HSS are named as the intermediate states, which are further divided into the hard intermediate states (HIMS) and the soft intermediate states (SIMS) based on the X-ray timing properties \citep{homan2005evolution, belloni2005evolution}. Typically, such transitions are often found to be accompanied by the changes in the types of LFQPOs, with type-C QPOs appearing mainly in the HIMS, whereas type-B and type-A QPOs appear only in the SIMS.

Despite the massive studies on BHTs, the evolution of the accretion disk/corona geometry is still under debate \citep{kara2019corona, you2021insight}. In the soft state, it is widely accepted that a geometrically thin and optically thick accretion disk has reached the innermost stable circular orbit (ISCO) of a BH. However, the disk/corona geometry in the LHS and HIMS is still an open question. The truncated disk geometry is most often proposed for the LHS and HIMS. Within the truncated disk model, the disk is assumed to be truncated at a radius that is larger than the ISCO and interior to which is the Comptonizing corona during the hard and intermediate states \citep{esin1997advection}. The transition from the hard state to the soft state may correspond to the decrease of the truncation radius. The truncated disc model has succeeded in explaining plenty of observed X-ray spectral and timing properties from BHXRBs, such as the hard–to–soft spectral transitions and the decreasing of the characteristic variability time-scale (see \citep{done2007modelling}, and references therein). Different from the truncated disk model, the lamppost model \citep{martocchia1996iron} assumes that a compact hard X-ray corona locates on the axis of the accretion disc. Under this scenario, the evolution of the source is associated with the vertical expansion or contract of the corona \citep{kara2019corona, buisson2019maxi, wang2021disk}. 

Another open question is the dynamical/radiative origin of LFQPOs observed in the LHS and HIMS. Several models have been proposed to explain the dynamical origin of LFQPOs considering either the instability in the accretion flow or a geometric effect of the accretion flow, among which the most promising model is the Lense-Thirring (L-T) precession model that assumes that LFQPOs are generated by the relativistic precession of an inner hot accretion flow \citep{ingram2009low, you2018x, you2020x} or a small-scale jet \citep{ma2021discovery}. Unless otherwise noted, the L-T precession mentioned in the rest of this paper all refers to the former one. In the L-T precession model, the precession frequency is set by parameters including the inner radius of a truncated accretion disk.  As the source spectra softens, the inner disk radius decreases and QPO frequency increases.
For the radiative origin of LFQPOs, \citet{2015MNRAS.446.3516I} reconstructed the QPO-phase dependent waveforms considering the rms and lags of the QPOs. This gave a description of the iron line energy shift at different QPO phases in GRS~1915+105. A time-dependent Comptonization model, \texttt{vKompth}, was proposed by~\citet{2020MNRAS.492.1399K} and~\citet{2022MNRAS.515.2099B}, to explain the energy dependent rms and lag spectra of the QPOs and measure the corona geometry~\citep{2021MNRAS.503.5522K,2022NatAs...6..577M}. From the measurements of this model, the corona geometry of BHTs can be slab-like or jet-like and connected to the jet behavior during the HIMS-to-SIMS transition in MAXI~J1535$-$571~\citep{2022MNRAS.512.2686Z,2023MNRAS.520.5144Z,rawat2023comptonizing}, MAXI~J1348-630~\citep{2021MNRAS.501.3173G}, and GX~339$-$4~\citep{2023MNRAS.519.1336P}.

In general, one way to study the geometry of the inner accretion flow is through the reflection spectrum. The hard photons from the corona could irradiate the accretion disk and are further reprocessed to produce a reflection component on the spectrum, i.e., the relativistic broadened Fe-K$\alpha$ emission line and the Compton hump component \citep{fabian2000broad,miller2007relativistic}. Plenty of efforts have been made by fitting the time-averaged reflection spectrum to estimate the BH spin, accretion disk inclination, and other characteristics. 

Another way to study the geometry of the inner accretion flow is to analyze the reverberation lags. Since the reflected disk photons have to travel a longer distance to the observer than the direct coronal photons, there will be a time delay between the two components, known as reverberation lags \citep[e.g.,][]{uttley2014x}. By applying the Fourier timing method, we are able to measure the time lags between these two components \citep[e.g.,][]{nowak1999rossi,uttley2014x}. Subsequently, through the measured reverberation lags, we can estimate the distance between the illuminated and reflected regions \citep{de2015tracing,de2021inner}. Combining the spectral analysis and reverberation lags study together gives us a better understanding of the accretion flow geometry.

X-ray reverberation lags are usually observed in radio-quiet active galactic nuclei (AGN) but rarely in BHXRBs. The time-scale of the lags scales linearly with the black hole mass. Since the mass of BHXRBs is much smaller than that of AGNs, the light travel time corresponding to one $R_{\rm{g}}$ is very short. The signal-to-noise ratio will be significantly reduced due to the small number of photons detected during the light travel time. Thus, reverberation lag detection is difficult in BHXRBs. The first detection of thermal X-ray reverberation lags in BHXRBs is in GX 339--4 \citep{uttley2011causal}. Further work by \citet{de2015tracing} showed that the reverberation lag of GX 339--4 decreases with increasing source luminosity and disk-fraction, which possibly supports a truncated disk geometry. To date, reverberation lags have been detected in several BHXRBs \citep{de2016reverberation, de2017evolution, kara2019corona, de2021inner, wang2020relativistic, wang2021disk, wang2022nicer}. 

MAXI J1535--571 was discovered as a new uncatalogued hard X-ray transient located near the Galactic plane by Monitor of All-Sky X-ray Image (\emph{MAXI}) on September 02, 2017 \citep{negoro2017maxi}. Follow-up observations were made by \emph{Swift}/BAT, \emph{INTEGRAL}, \emph{Insight}-HXMT, \emph{NuSTAR}, and \emph{NICER}. Due to its behavior observed in X-Ray and Radio bands, MAXI J1535--571 is classified as a bright BHXRB candidate \citep{negoro2017further}. LFQPOs have been detected by \emph{Insight}-HXMT and \emph{NICER} in the LHS, HIMS, and SIMS \citep{huang2018insight, stiele2018spectral}. The spectral analysis of the \emph{NuSTAR} observations gives a black hole spin a $>$ 0.84 and an inclination angle i = ${57_{-2}^{+1}}^\circ$  \citep{xu2018reflection}. \citet{chauhan2019h} estimated a distance of $4.0\pm0.2$ kpc for the source by studying the HI absorption from gas clouds along the line-of-sight.

In this paper, we study the accretion flow geometry of MAXI J1535--571 by applying two independent methods: broadband energy spectrum fitting and reverberation lags analysis. The data analyzed in this paper are taken from \emph{Insight}-HXMT and \emph{NICER}, covering both the hard and intermediate states during the 2017 outburst. Considering \emph{NICER}'s high time resolution and large area in the soft energy band (0.2-10 keV), we mainly use \emph{NICER} data for timing analysis. The \emph{NICER} observations also cover the entire transition states. On the other hand, considering \emph{NICER}'s narrower energy band  and its calibration uncertainty below 3 keV \citep{miller2018nicer}, we think that \emph{Insight}-HXMT has more advantages in broad band spectral fitting, especially above 20 keV. Therefore, we choose \emph{Insight}-HXMT data for energy spectrum fitting.

This paper is organized as follows. Section~\ref{sec2} describes the observations and data reduction. Section~\ref{sec3} provides the time lag analysis with \emph{NICER} data. The details of broad energy band spectrum fitting with \emph{Insight}-HXMT data are described in Section~\ref{sec4}. Discussions and conclusions are presented in Section~\ref{sec5}.

\section{DATA REDUCTION} \label{sec2}

The data set analyzed in this paper includes 29 \emph{NICER} and 28 \emph{Insight}-HXMT observations carried out between September 12th and October 11th, 2017. The selected \emph{NICER} ObsIDs are from 1050360104 to 1130360114 and \emph{Insight}-HXMT ObsIDs are from P0114535001 to P0114535009. Table~\ref{tableA1} and Table~\ref{tableA2} list the log of the observations. The \emph{NICER} and \emph{Insight}-HXMT hardness-intensity diagrams (HIDs) and hardness-rms diagrams (HRDs) are shown in Figure~\ref{hid}. The accretion state classifications of \emph{Insight}-HXMT observations is taken from \citet{huang2018insight}. It can be seen from the HRD that the SIMS locates in the lower left of the diagram due to low variability and hardness. However, \emph{Insight}-HXMT only covers observations before MJD 58020. Through \emph{NICER} observations, we can see that after MJD 58027 the data points return to the right top of the HRD. Meanwhile, according to the timing analysis of \citet{stiele2018spectral}, the QPO type changes from A to C, and the associated noise component changes from red noise to flat-top noise, all of which indicate that the source has returned to the HIMS \citep{belloni2010states}. As one of the signs of state transition, type-B QPOs were detected by \emph{Insight}-HXMT at MJD 58016 \citep{huang2018insight}, but were missed by \emph{NICER} due to the lack of observations \citep{stevens2018nicer,stiele2018spectral}.

The \emph{NICER} data are processed with the NICERDAS tools
in HEASOFT v.6.27.2 and CALDB v.20200722. The data are screened using the standard calibration tool NICERCAL and screening tool NIMAKETIME. We select events that are caught less than $54^{''}$ offset in pointing, more than $40^{\circ}$ away from the bright Earth limb, more than $30^{\circ}$ away from the dark Earth limb, outside the South Atlantic Anomaly (SAA), not flagged as “overshoot” or “undershoot” resets (EVENT FLAGS=bxxxx00), and triggered the slow signal chain (EVENT FLAGS = bx1x000). A “trumpet” filter is also applied to eliminate known background events \citep{bogdanov2019constraining}. 

The Hard X-ray Modulation Telescope, known as \emph{Insight}-HXMT \citep{zhang2014introduction}, consists of three groups of instruments: the high-energy X-ray telescope (HE, 20--250 keV, 5,100 cm$^2$), the medium-energy X-ray telescope (ME, 5--30 keV, 952 cm$^2$), and the low-energy X-ray telescope (LE, 1--15 keV, 384 cm$^2$). HE contains 18 cylindrical NaI(Tl)/CsI(Na) phoswich detectors; ME is composed of 1728 Si-PIN detectors; and LE uses Swept Charge Device (SCD). There are three types of Field of View (FoV): 1$^\circ$ × 6$^\circ$ (i.e., the small FoV), 6$^\circ$ × 6$^\circ$ (i.e., the large FoV), and the blind FoV used to estimate the particle induced instrumental background. More details about \emph{Insight}–HXMT can be found in \citet{zhang2020overview}.

The \emph{Insight}-HXMT data are processed with \emph{Insight}-HXMT Data Analysis Software (HXMTDAS) version 2.03. The data are filtered using the criteria recommended by the \emph{Insight}-HXMT team: the pointing offset angle is smaller than 0.04$^\circ$; the elevation angle is larger than 10$^\circ$; the value of the geomagnetic cutoff rigidity is larger than 8; data are used at least 300 s before and after the South Atlantic Anomaly (SAA) passage. The energy bands chosen for energy spectral analysis are 2-10\ keV (LE), 10-27\ keV (ME), and 27-80\ keV (HE). The XSPEC v12.11.0c software package \citep{arnaud1996xspec} is used to perform spectral fitting. All parameter uncertainties are estimated at the 90\% confidence level.

For \emph{NICER} data, we generate the cross-spectrum using standard techniques \citep{nowak1999rossi, uttley2014x} to compute the X-ray lags as a function of Fourier-frequency. The energy bands we select to compute the spectra are 0.5-2.5\ keV (soft band) and 3-5\ keV (hard band), in which the soft band is dominated by the reflected disk photons and the hard band is dominated by the coronal photons. To avoid interference from the iron K lines and iron edge, we select the energy bands below 5 keV. A positive (hard) lag means that the hard photons lag behind the soft ones. It is worth mentioning that 0.5-1 keV is often used as the soft band in the study of the reverberation lags, as the soft excess is more significant below 1 keV \citep{kara2019corona,de2021inner}. However, MAXI J1535-571 has a relatively higher interstellar absorption of $N_{\rm{H}} = (3 - 8)\times10^{22} cm^{-2}$ \citep{tao2018swift,xu2018reflection,kong2020joint}. The high absorption significantly diminishes the photons at soft X-rays ($<$ 1 keV), which consequently leads the quality of the lag spectra too poor to confidently measure the reverberation lag and its characteristic frequency in the soft band of 0.5–1 keV. Extending the soft energy band to 0.5-2.5 keV can largely improve the signal-to-noise ratio, while according to \citet{uttley2014x}, the soft excess generally exists below 3 keV. Moreover, since photons above 1 keV are less affected by absorption, the soft excess within 1-2.5 keV in MAXI J1535-571 could be more significant than within 0.5-1 keV.

\section{TIMING ANALYSIS} \label{sec3}

\begin{figure}
	\includegraphics[width=\columnwidth]{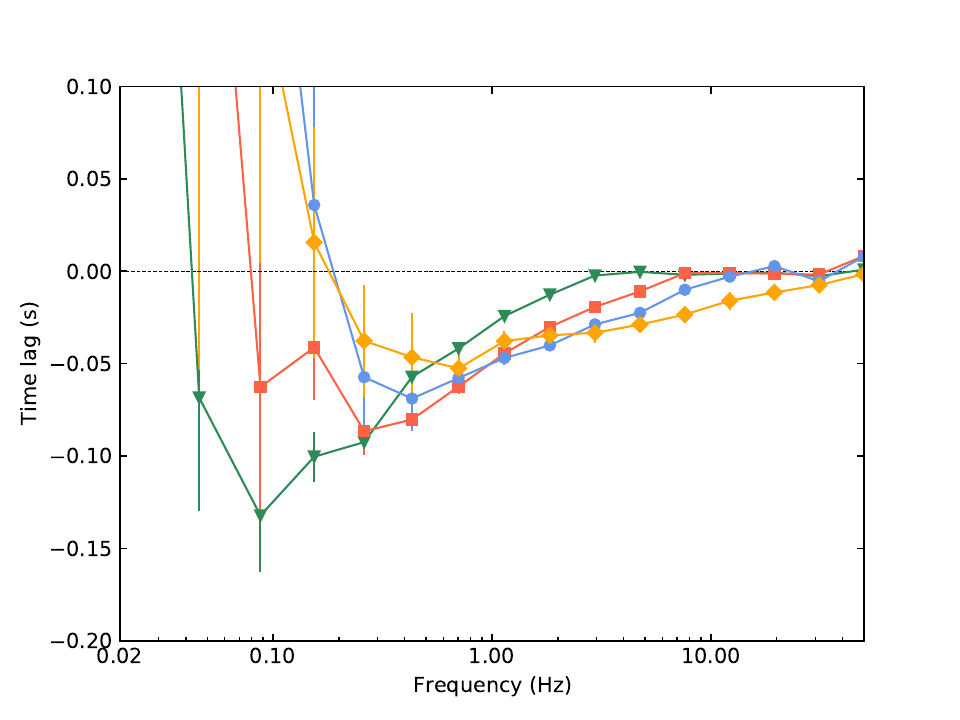}
    \caption{The 0.5--2.5 keV vs. 3--5 keV lag-frequency spectra of some of the analysed \emph{NICER} observations of MAXI J1535--571. A negative lag suggests that the soft band lags behind the hard ones. Green triangles, red squares, blue dots, and orange diamonds represent ObsID 1050360106, 1050360110, 1130360103 and the combined SIMS observations, respectively.}
    \label{ccf}
\end{figure}

\begin{figure}
	\includegraphics[width=\columnwidth]{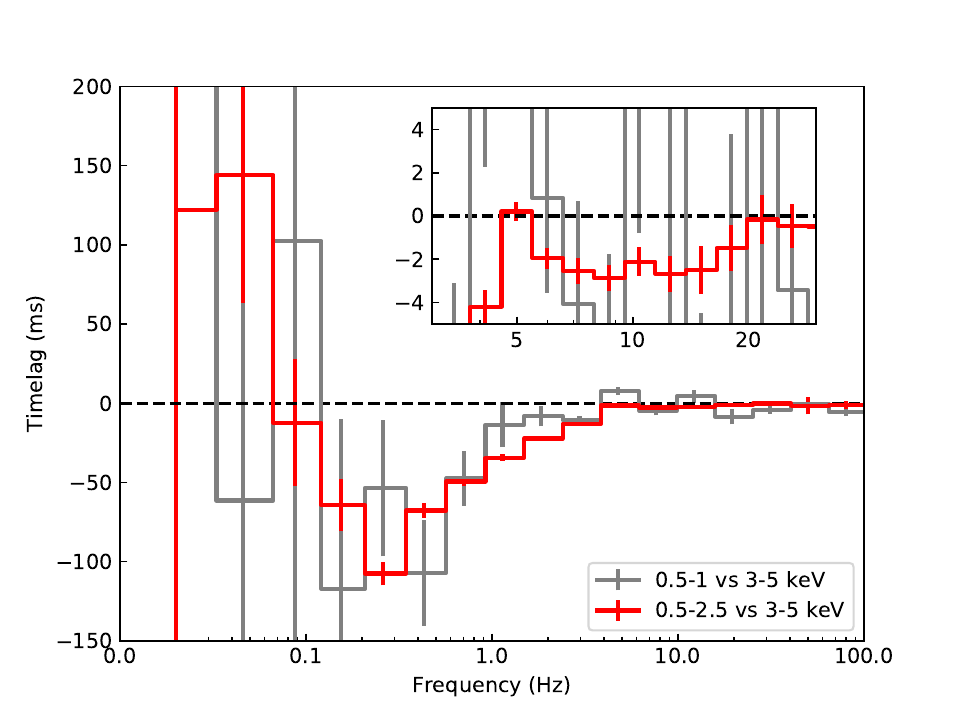}
    \caption{The 0.5--2.5 keV vs. 3--5 keV and 0.5--1 keV vs. 3--5 keV lag-frequency spectras of \emph{NICER} observation 1050360104. }
    \label{ccf1}
\end{figure}

\begin{figure*}
	\centering\includegraphics[width=\columnwidth]{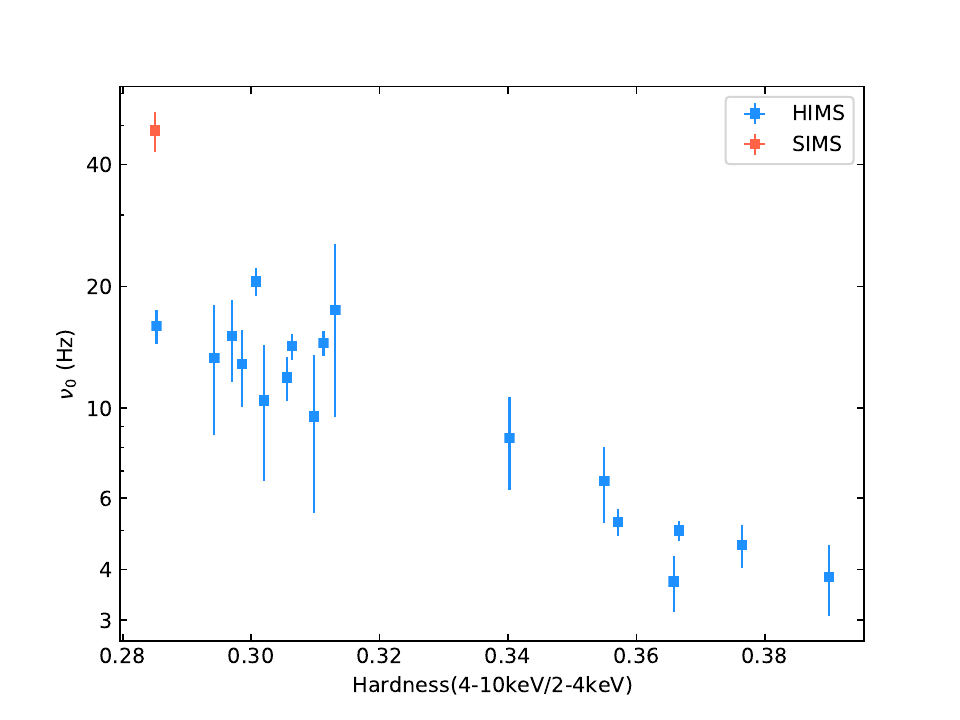}
    \centering\includegraphics[width=\columnwidth]{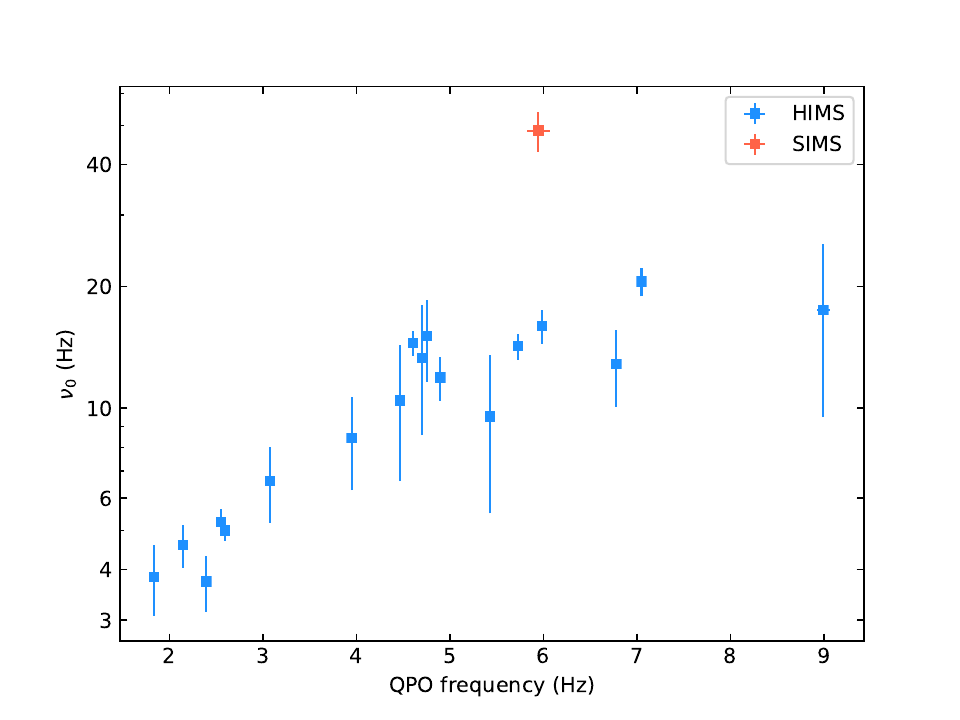}
    \caption{The characteristic frequency $\nu_0$ as a function of spectral hardness (left panel) and QPO frequency (right panel).}
    \label{v0-relation}
\end{figure*}

A number of examples of lag-frequency spectra are shown in Figure~\ref{ccf}. The evolution of lag-frequency spectra is not significant in the SIMS. Due to the low source variability level, the observations from SIMS are combined together.

At low frequencies, we observed hard (positive) X-ray lags in all the analysed observations. These lags are commonly observed in BHXRBs \citep[e.g.,][]{miyamoto1988delayed,nowak1999rossi, pottschmidt2000temporal}. Previous studies suggest that a power-law model, with index of $\sim$ $-$ 0.7, is able to qualitatively describe the underlying decreasing trend of hard lags as a function of frequency in BHXRBs \citep{de2017evolution}. The low-frequency hard lags are usually interpreted as the inward propagation of fluctuations in the disc mass accretion rate \citep{kotov2001x, arevalo2006investigating, ingram2013exact}.

At high frequencies, soft (negative) X-ray lags are clearly observed in all the observations. The soft lags evolve significantly in both frequency and amplitude throughout the outburst. The frequency of the soft lags increases with the decreasing hardness ratio, while the amplitude of them decreases with the decreasing hardness ratio. These high-frequency soft X-ray lags are usually introduced by reverberation caused by the light-crossing time delay between the continuum and reflected emission.

However, since the lag is measured between two energy bands that cover both the irradiation and reflection components, the observed soft lags will consequently be diluted, which is known as the dilution effect \citep{uttley2014x}. Because of the dilution effect, the amplitude of soft lag cannot accurately reflect the intrinsic reverberation lags. Therefore, we adopt a method introduced by \citet{de2021inner} by taking the frequency at which the soft lag first turns to zero (hereafter $\nu_0$) as the intrinsic time scale of reverberation lags. In this way, the dilution effects can be avoided.

In Figure~\ref{ccf1}, the soft lag approaches zero at $\sim$5 Hz and then rapidly turns negative when frequency increases. Notably, for a mildly rebinned lag frequency spectrum, only one or two bins cross the zero-lag may be just a statistical fluctuation. In addition, as the critical frequency of phase wrapping, the lag above $\nu_0$ is generally positive, which is different from what we have observed. However, in real cases, the phase wrapping is highly affected by the response function and dilution effect. For certain response functions, phase wrapping does not necessarily lead to positive lags \citep[see Fig 21 in ][]{uttley2014x}. On the contrary, the lags may change from zero to negative values above $\nu_0$. In other words, the lag curve does not have to cross the zero-lag line and $\nu_0$ is more like an inflection point, which is the case observed in MAXI J1535-571. Since there could be different response functions for different sources, we would see different lag curves for different sources. In addition, the inflection point around $\sim$5 Hz doesn’t change with the dilution effect, which proves one of the main properties of the characteristic frequency $\nu_0$.

Considering that the positive lags near $\nu_0$ is very small, if we measure $\nu_0$ directly, it will be highly affected by the rebin factor. In order to quantitatively measure $\nu_0$ in a model-independent way, we use a logarithm function $f(x)=a+blnx$ to fit parts of the lag-frequency spectra near the high frequency zero point. And to reduce the bias introduced by fitting, we only selected four bins to fit. Some fitting examples are shown in Fig~\ref{ccf_all}. We further plot the $\nu_0$ as a function of hardness ratio and QPO frequency. The QPO frequency values are taken from \citet{stiele2018spectral}. As shown in Figure~\ref{v0-relation}, during the HIMS (blue dots), $\nu_0$ is inversely correlated with the hardness ratio, while positively correlated with type-C QPO frequency. During the SIMS, $\nu_0$ increases dramatically, reaching two or three times higher than in the HIMS for both plots. 

Since $\nu_0$ is used as the time scale of the reverberation mapping, it can qualitatively describe the distance between the corona and the disk. Hence, during the HIMS, the distance between the disk and corona decreases as the source softens. When the source enters the SIMS, the distance significantly decreases. These behaviours suggest a possible moving inward disc scenario in the truncated disk geometry or a decrease of the corona height in the lamppost geometry. In the former case, the inner disc probably reaches the ISCO in the SIMS. A more detailed discussion is given in Section~\ref{sec5}.

\section{SPECTRAL ANALYSIS} \label{sec4}

\begin{figure}
	\centering\includegraphics[width=\columnwidth]{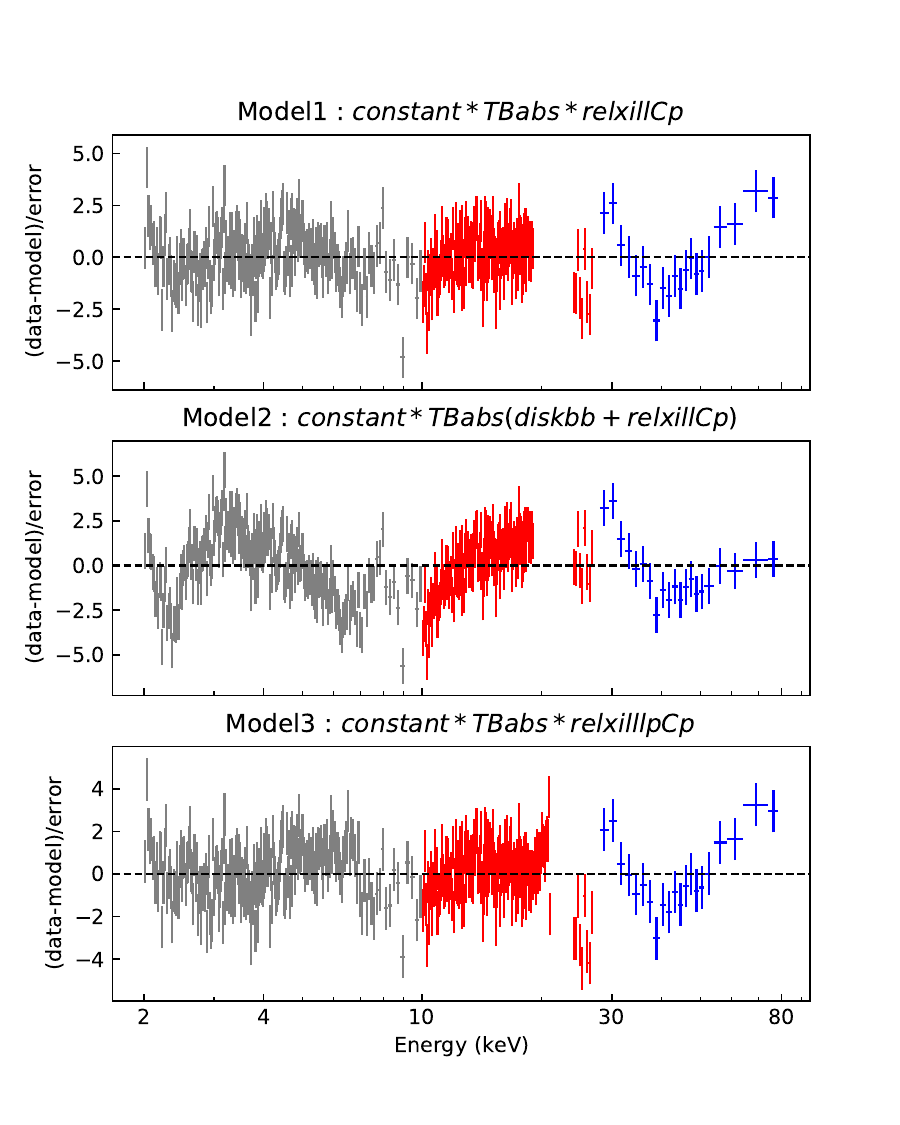}
    \caption{(Data-model)/error plots of the reflection modeling of the observation P011453500501. The gray points: LE (2 -- 10 keV); the red points: ME (10 -- 27 keV); the blue points: HE (27 -- 80 keV).}
    \label{ratio}
\end{figure}

\begin{figure*}
	\centering\includegraphics[width=\columnwidth]{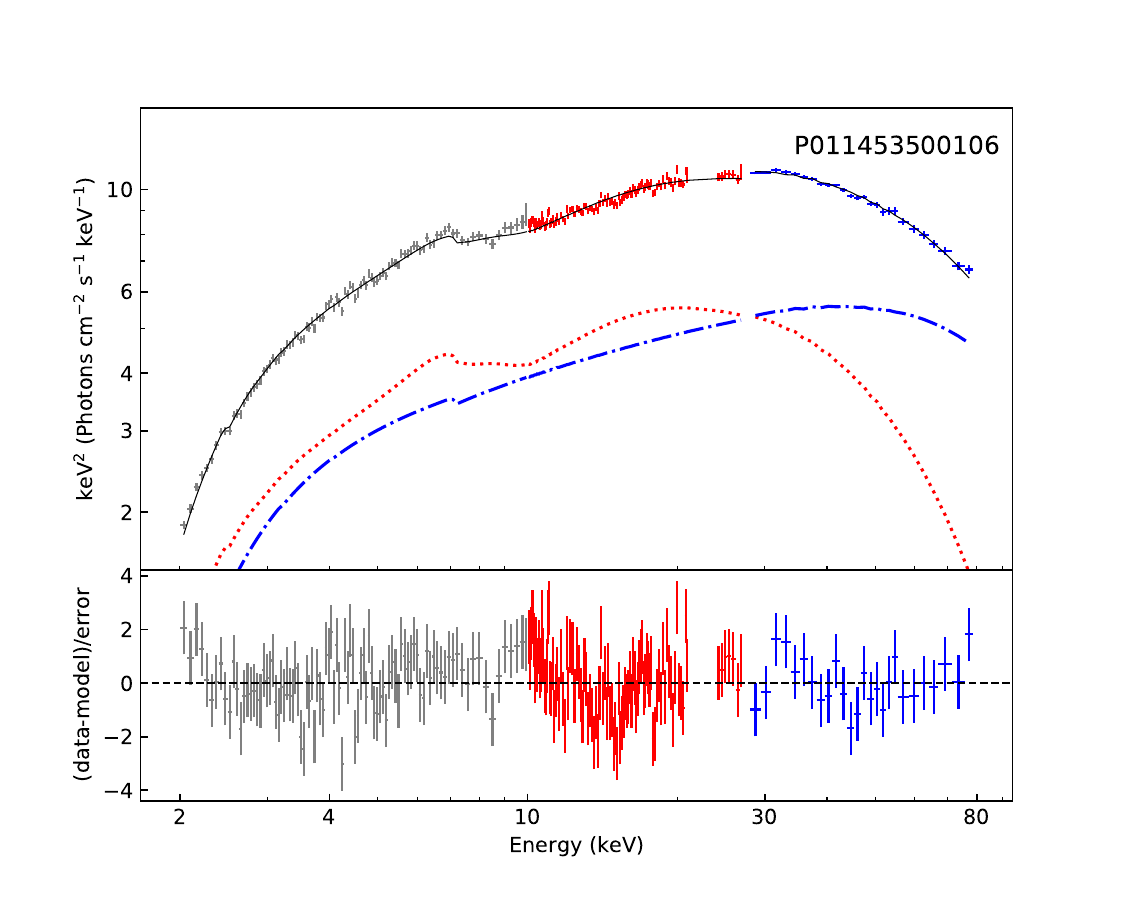}
	\centering\includegraphics[width=\columnwidth]{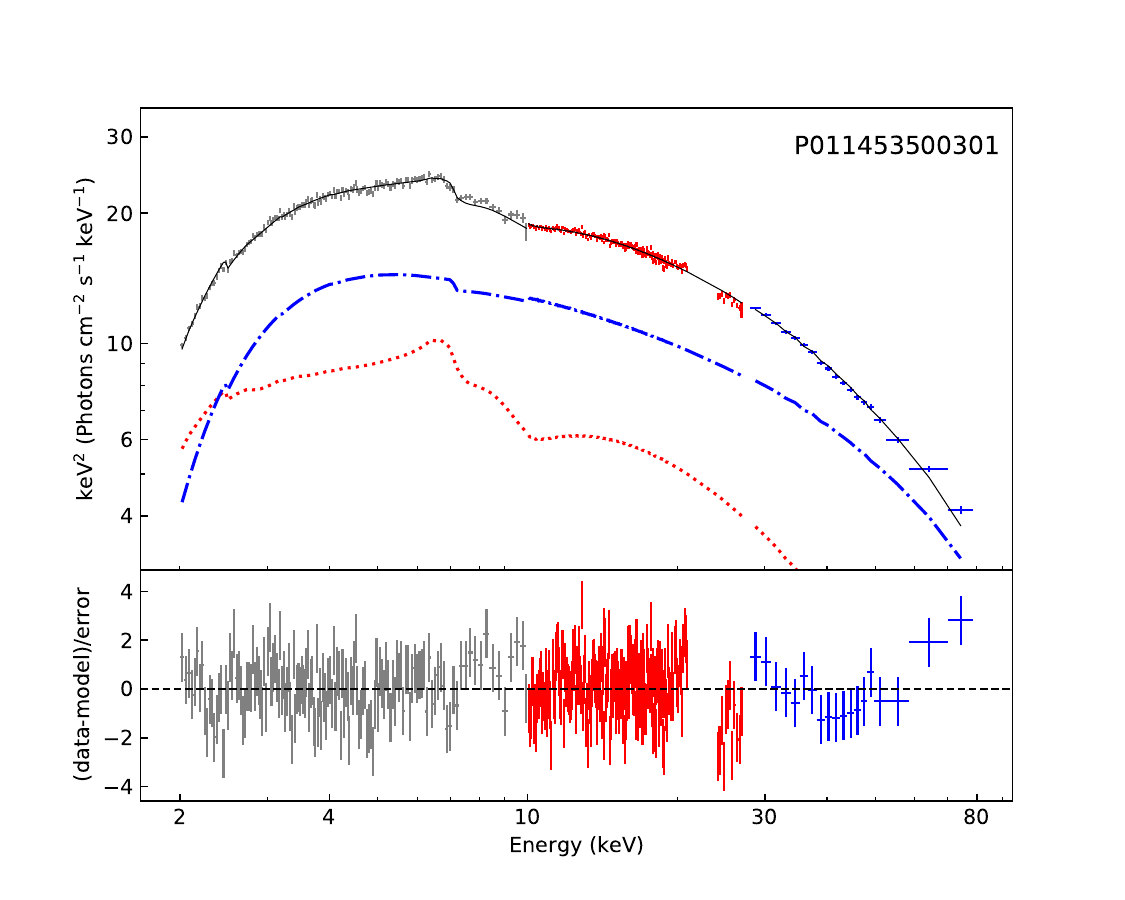}
	\centering\includegraphics[width=\columnwidth]{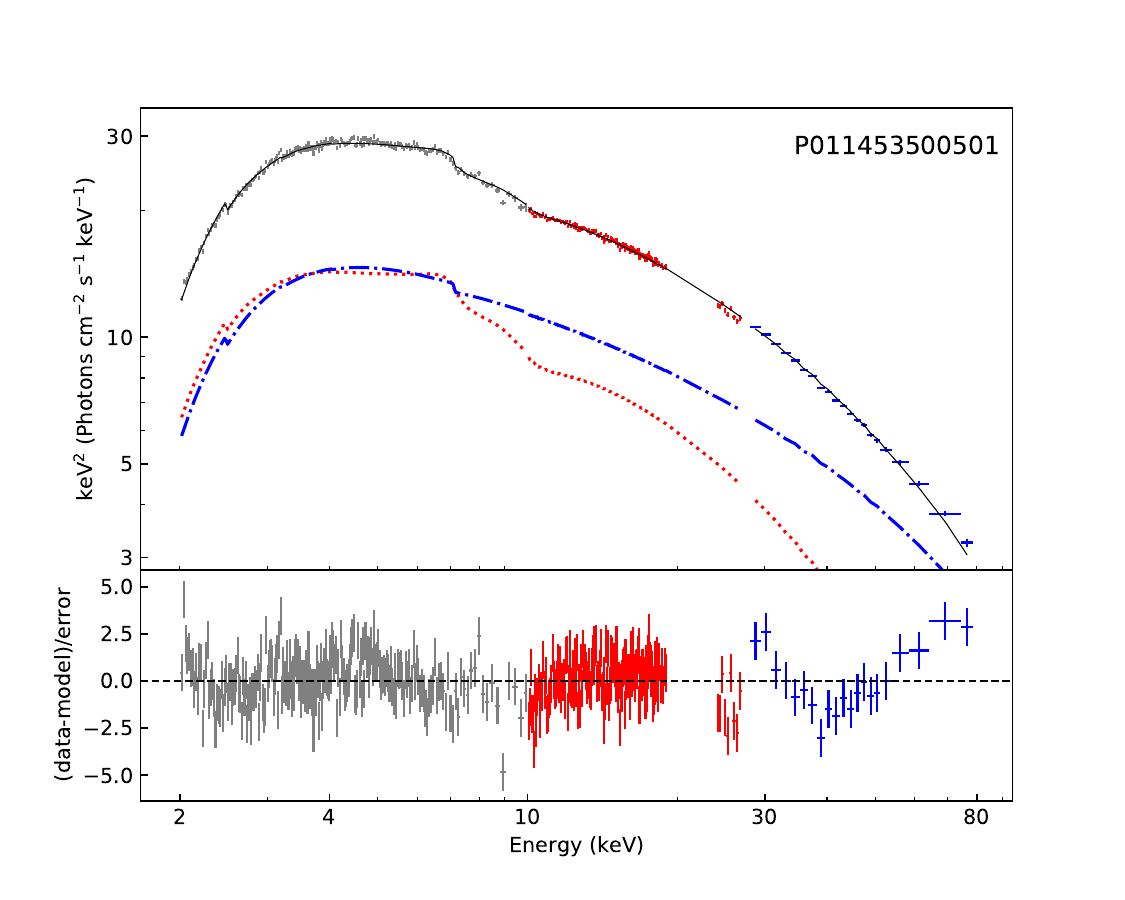}
	\centering\includegraphics[width=\columnwidth]{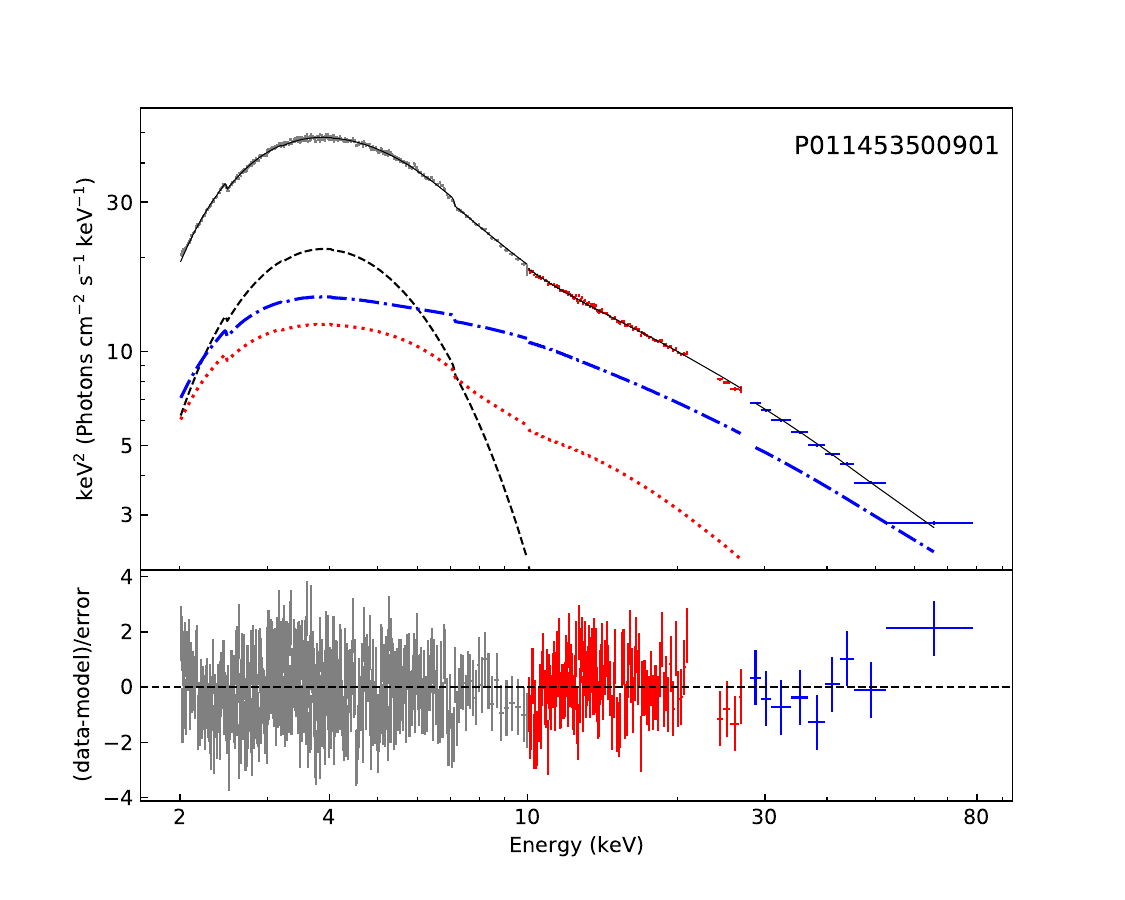}	
    \caption{Spectral fittings with \emph{Insight}-HXMT observations P011453500106, P011453500301, P011453500501 and P011453500901. The gray, red, and blue points are for LE, ME, and HE, respectively. The total model is shown in thick-solid line; the thermal emission \emph{(diskbb)} from the disc is shown in black dashed line; the Compotonization component \emph{nthcomp} is shown in blue dot-dashed line, and it is calculated internally by \emph{relxillcp}; the relativistic reflection component is shown in red dotted line.}
    \label{ratio1}
\end{figure*}

\renewcommand\arraystretch{1.2}
\begin{table*}[htbp]
    \begin{center}
     \caption{Best fitting parameters for model1 : $constant*TBabs*(relxillCp+diskbb)$}.
        \setlength{\tabcolsep}{1mm}{
	    \begin{tabular}{cccccccccccc}
		    \hline
		    $^a\rm{ObsID}$ & $N_{\rm{H}}$ & $T_{\rm{in}}$ & $N_{\rm{disk}}$ & $q$  & $R_{\rm{in}}$ & $\Gamma$ & log$\xi$ & $kT_{\rm{e}}$ & $R_{\rm{f}}$ & $N_{\rm{rel}}$ & $\chi^{2}_{\rm{red}}(d.o.f)$\\
		     & ($10^{22}{\rm{cm}}^{-2}$) & (keV) &($10^3)$ & & (ISCO) &  &  & (keV) &  & \\
		    \hline
            106 & $4.34_{-0.10}^{+0.05} $ & $-$ & $-$ &$2.2_{-1.0}^{+0.9} $ & $26_{-13}^{+20} $ & $1.75_{-0.01}^{+0.01} $ & $3.6_{-0.03}^{+0.02} $ & $28.2_{-1.1}^{+1.2} $ & $0.38_{-0.04}^{+0.04} $ & $0.05_{-0.01}^{+0.01} $ & 1.06(1206) \\
            145 & $4.96_{-0.02}^{+0.10} $ & $-$ & $-$ &$2.6_{-0.6}^{+0.5} $ & $38_{-15}^{+9} $ & $2.31_{-0.01}^{+0.01} $ & $3.87_{-0.07}^{+0.04} $ & $33.9_{-0.6}^{+2.4} $ & $0.24_{-0.02}^{+0.04} $ & $0.28_{-0.02}^{+0.02} $ &1.19(1174) \\
            301 & $4.75_{-0.03}^{+0.04} $ & $-$ & $-$ &$1.8_{-0.4}^{+0.4} $ & $35_{-19}^{+12} $ & $2.28_{-0.01}^{+0.01} $ & $3.87_{-0.04}^{+0.03} $ & $25.7_{-2.8}^{+1.8} $ & $0.20_{-0.03}^{+0.02} $ & $0.28_{-0.02}^{+0.01} $ &1.07(1204)\\
            401 & $4.74_{-0.02}^{+0.03} $ & $-$ & $-$ &$1.2_{-0.6}^{+0.8} $ & $34_{-11}^{+17} $ & $2.34_{-0.01}^{+0.01} $ & $4.26_{-0.04}^{+0.02} $ & $34.8_{-0.8}^{+1.6} $ & $0.32_{-0.02}^{+0.02} $ & $0.26_{-0.01}^{+0.01} $ &1.08(1206)\\
            501 & $5.04_{-0.01}^{+0.07} $ & $-$ & $-$ &$1.4_{-0.3}^{+0.4} $ & $8.8_{-3.1}^{+3.4} $ & $2.42_{-0.01}^{+0.01} $ & $4.43_{-0.02}^{+0.06} $ & $43.6_{-0.7}^{+2.9} $ & $0.35_{-0.01}^{+0.03} $ & $0.34_{-0.01}^{+0.01} $ &1.07(1172)\\
            601 & $5.05_{-0.14}^{+0.08} $ & $-$ & $-$ &$1.4_{-0.3}^{+0.2} $ & $10.3_{-2.2}^{+3.0} $ & $2.43_{-0.01}^{+0.01} $ & $4.27_{-0.01}^{+0.06} $ & $42.9_{-0.5}^{+7.8} $ & $0.37_{-0.05}^{+0.06} $ & $0.35_{-0.03}^{+0.03} $ &1.06(1150)\\
            901 & $5.10_{-0.07}^{+0.11} $ & $1.20_{-0.03}^{+0.01}$ & $1.93_{-0.05}^{+0.04}$ &$6.2_{-0.2}^{+0.5} $ & $1.14_{-0.08}^{+0.14} $ & $2.64_{-0.01}^{+0.02} $ & $4.34_{-0.16}^{+0.31} $ & $289_{-25}^{+30} $ & $0.50_{-0.02}^{+0.03} $ & $0.58_{-0.04}^{+0.02} $ &1.15(1204)\\
            906 & $4.62_{-0.05}^{+0.08} $ & $1.15_{-0.01}^{+0.01} $ & $4.38_{-0.06}^{+0.08} $ &$7.7_{-0.3}^{+0.4} $ & $1.61_{-0.21}^{+0.16} $ & $2.76_{-0.01}^{+0.01} $ & $3.71_{-0.07}^{+0.23} $ & $400_{-15}^{+0} $ & $0.55_{-0.01}^{+0.02} $ & $0.34_{-0.02}^{+0.02} $ &1.13(1204)\\
            912 & $4.64_{-0.06}^{+0.12} $ & $1.16_{-0.02}^{+0.01}$ & $4.22_{-0.04}^{+0.06}$ &$7.3_{-0.3}^{+0.3} $ & $1.58_{-0.14}^{+0.08} $ & $2.71_{-0.01}^{+0.01} $ & $4.04_{-0.14}^{+0.09} $ & $400_{-13}^{+0} $ & $0.55_{-0.02}^{+0.04} $ & $0.30_{-0.03}^{+0.01} $ &1.06(1204)\\
		    \hline
	    \end{tabular}
        }
        \label{table1}
    \end{center}
    \begin{tablenotes}
    \footnotesize
    \item{a} P011453500XXX
    \end{tablenotes}
\end{table*}

In order to further study the reflection characteristics of the source, we use the relativistic reflection model to fit the \emph{Insight}-HXMT energy spectra of MAXI J1535--571. For all the spectra, a neutral Galactic absorption component, modeled with $TBabs$ is added \citep{wilms2000absorption}. We adopt the abundances in \citet{wilms2000absorption} as appropriate for absorption by the Galactic interstellar medium and adopt the recommended cross-sections of \citet{verner1996atomic}. Fluorescence
lines due to the photoelectric effect of electrons in K-shell of silver are detected by the Si-PIN detectors of ME, which contributes to the spectra at 21-24 keV. Therefore, the data points in 21-24 keV are ignored. 

We select a HIMS observation and fit its spectra with model $TBabs*relxillCp$ (Model 1). The relativistic reflection model $relxillCp$ \footnote{The version of $relxill$ model used in this work is 1.4.0.} contains both the emission component of the corona and the reflection component of the disk \citep{dauser2014role}. The emission component is described by Comptonization model $nthcomp$. Previous measurements have shown that MAXI J1535--571 has a high black hole spin of $>$ 0.84 in \citet{xu2018reflection}, and 0.994(2) in \citet{miller2018nicer}. In order to reduce the parameter space, we fix the spin at its maximum value of 0.998, considering that the adoption of other spin values does not change our main conclusions from our attempt to fit the data. Our data cannot simultaneously constrain the $q_{\rm{1}}$, $q_{\rm{2}}$ and $R_{\rm{br}}$. In order to allow the inner disk radius to fit to any physically allowed value, we use a simple powerlaw to describe the emissivity profile by linking $q_{\rm{1}}$ to $q_{\rm{2}}$. If assume the Newtonian case, i.e., $q_{\rm{2}}$ is fixed at 3, we get much worse fits than let $q_{\rm{2}}$ vary freely. The $\Delta\chi^{2}$ between $q_{\rm{2}}$ fixed at 3 and $q_{\rm{2}}$ free to vary are 67.24, 52.61, 42.56 and 36.68 for Obs 106, 301, 501 and 901, respectively. An emissivity profile with $q_{\rm{2}}=3$ is usually considered as a standard scenario, under the assumption that the intensity of the hard radiation scattered back on to the disk by the corona is proportional to the local disk emissivity \citep{shakura1973black,dauser2013irradiation}. However, non-thermal coronal emission does not necessarily need to behave in the same way as the thermal dissipation of the disk. The interaction between the disc and the corona is more complicated, including the radiation and magnetic processes \citep{haardt1991two,czerny2004flare,goosmann2006magnetic,rozanska2011iron}. $N_{\rm{H}}$ is set free since it's affected by the environment around the compact star, such as the accretion disk, interstellar gas, and outflow matter.

The HIMS X-ray spectra can be well fitted by Model 1 with a reduced chi-square $\chi_{\rm{\nu}}^2=1254/1172=1.07$. The residual is shown in Figure~\ref{ratio}. Adding an extra $diskbb$ component (Model 2) can neither improve the goodness-of-fit nor constrain the disk parameters. If we fix the disk temperature and norm to the values suggested by \citet{kong2020joint}, the reduced chi-square $\chi_{\rm{\nu}}^2$ increases to 1899/1231=1.54 (see Figure~\ref{ratio}). Obviously, a disk component is not required for HIMS, we only add a $diskbb$ component for the SIMS observations.

We also try to fit the spectrum with a lamppost model, $TBabs*relxilllpCp$ (Model 3). The model $relxilllpCp$ is also a member of $relxill$ family \citep{dauser2014role}. The parameter $fixReflFrac$ is fixed to 1, so that the reflection fraction can be self-consistently calculated, according to the configurations of other parameters, e.g., the BH spin, inner disk radius, and the height of the lamppost source \citep{dauser2014role}. Model 3 also gives a good fit with a reduced chi-square of 1419/1207=1.18 (see Figure~\ref{ratio}). However, by plotting the contour, we find that the corona height $h$ is degenerated with the inner radius of the disk $R_{\rm{in}}$ (see Figure~\ref{fig12}) and we cannot determine $h$ and $R_{\rm{in}}$ independently from this model. Note that the degeneracy between $R_{\rm{in}}$ and $h$ is positively correlated here, which is in contrast to the regular case. We find it could be caused by the degeneracy among the other parameters. In Table~\ref{tableA3}, we give the best-fitting values of the parameters from HXMT observations using Model 3, from which it suggests that the evolution of $h$ and $R_{\rm{in}}$ is substantially affected by the degeneracy. Nevertheless, at the 90\% confidence level, our results strongly suggest that the accretion disk is truncated in the HIMS. Since we are mainly concerned with the detailed geometric evolution of the source, Model 1 is finally adopted. Although Model 1 has no assumption about the coronal geometry, it can give information about the inner radius of the disk.

\begin{figure*}
	\centering\includegraphics[width=14cm]{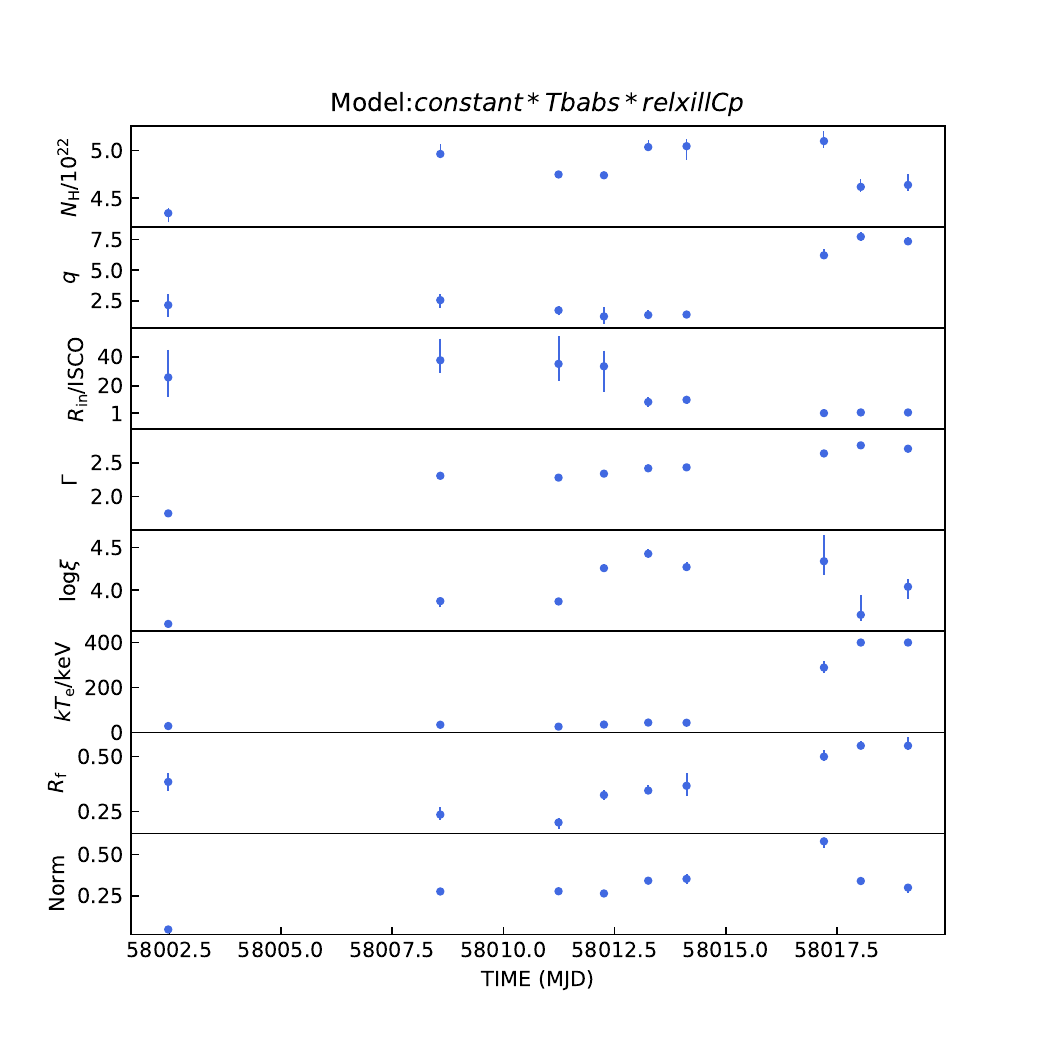}
    \caption{The evolution of Insight-HXMT spectral parameters. The parameters are listed in Table~\ref{table1}.}
    \label{spec_par}
\end{figure*}

In Figure~\ref{ratio1}, we show the spectra taken from the LHS, HIMS, and SIMS. The deep of the residuals around the Compton hump region is due to the calibration of the high energy detector. Due to the low S/N, the energy spectra of some observations could not constrain the parameters well, hence are removed from our analysis. The fitting parameters for all the rest observations are shown in Table~\ref{table1}. The evolution of the spectral parameters are given in Figure~\ref{spec_par}. The model gives an inclination angle i of ${51.2_{-1.2}^{+0.6}}^{\circ}$ and an iron abundance AFe of $0.52_{-0.01}^{+0.03}$ (in solar units). The evolution trend of the photon index $\Gamma$ is consistent with the previous \emph{Insight}-HXMT results given by \citet{kong2020joint}, and \emph{Swift} results given by \citet{tao2018swift}. The ionization parameter log$\xi$ varies between 3.6 and 4.4.

The column density $N_{\rm{H}}$ (in units of $10^{22}{\rm{cm}}^{-2}$) has a value higher than the Galactic absorption column density $N_{\rm{H}} = 1.5 \times 10^{22} \rm{cm^{-2}}$ \citep{kalberla2005leiden}, evolving in the range of 4.3 and 5.1. High $N_{\rm{H}}$ and its variation of nearly 20\% during the outburst have also been observed in previous studies \citep{xu2018reflection, tao2018swift, kong2020joint}. In \citet{tao2018swift}, they propose that when the accretion rate increases, the outflow from the disk will lead to the observed $N_{\rm{H}}$ increase.

The inner disk radius $R_{\rm{in}}$ is truncated at about 25 $R_{\rm{ISCO}}$ at the beginning of the LHS, then steadily decreases to 10 $R_{\rm{ISCO}}$ at the end of the HIMS. After the source entered the SIMS, $R_{\rm{in}}$ is close to the ISCO, as shown in Figure~\ref{spec_par}. These results suggest that MAXI J1535--571 most probably has a truncated disk. While the inner edge of the truncated disk moves inward towards the BH, the relative distance between the corona and the disk decreases, which is also supported by the results of reverberation lag analysis in Section~\ref{sec3}.

\section{SUMMARY AND DISCUSSION} \label{sec5}

In this paper, we have performed a detailed spectral-timing analysis of the BHXRB MAXI J1535--571 during its 2017 outburst, using observations from \emph{NICER} and \emph{Insight}-HXMT. We find that the geometry of the inner accretion flow has evolved significantly from the LHS to the SIMS. In particular, the characteristic frequency $\nu_0$ of reverberation lags increases during the HIMS, as shown in Figure~\ref{v0-relation}, suggesting that the relative distance between the disk and the corona decreases when the spectra softens. We further studied the reflection characteristics in a broad energy band with \emph{Insight}-HXMT data. We propose that the disk is truncated in the hard state and reach the ISCO in the soft intermediate state.

\begin{figure}
	\centering\includegraphics[width=\columnwidth]{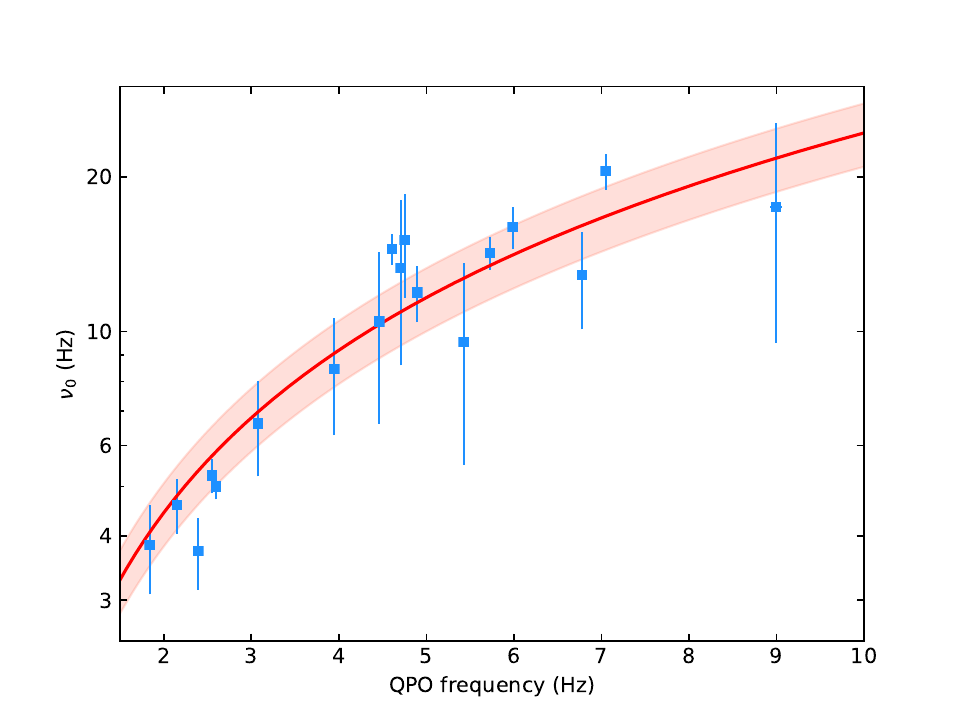}
    \caption{$\nu_0$ as a function of QPO frequency. The red line shows the best fit of the L-T precession model (95\% confidence level).}
    \label{LT}
\end{figure}

\begin{figure}
	\centering\includegraphics[width=\columnwidth]{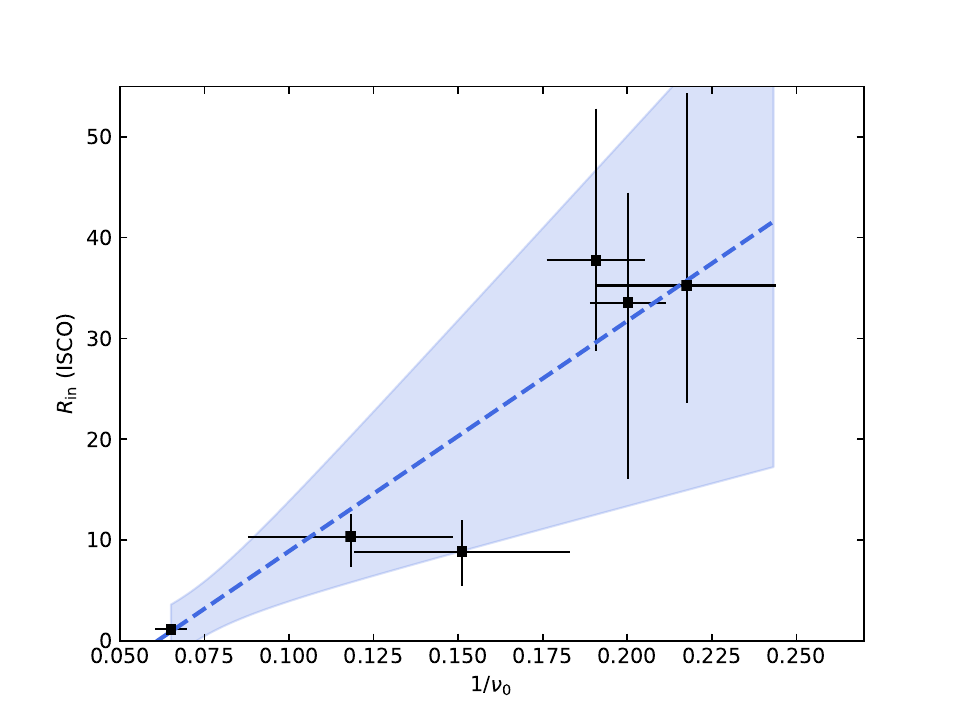}
    \caption{The reverberation time-scale as a function of the inner disk radius $R_{\rm{in}}$. We select the quasi-simultaneous observations of \emph{NICER} and \emph{Insight}-HXMT ( $\pm$ 1 day). The blue dashed line shows the best linear fit (95\% confidence level). }
    \label{Rin_v0}
\end{figure}

During the HIMS, the reverberation characteristic frequency $\nu_0$ shows a positive correlation with the type-C QPOs frequency, as shown in Figure~\ref{v0-relation}. In the L-T precession model, when the inner disk radius moves inwards, the QPO frequency increases, meanwhile the relative distance between the disk and the corona decreases, leading to a decrease of the characteristic frequency $\nu_0$ of reverberation lags.

According to \citet{ingram2014solutions}, the QPO frequency in the L-T precession model can be calculated by the following formula: 

\begin{equation}
    \frac{f_{\rm{QPO}}}{f_{\rm{K}}^*} =\left(1-\sqrt{1-\sqrt{2}ar^{3/2}+0.75a^2r^2}\right)
	\label{eq1}
\end{equation}

\begin{equation}
    f_{\rm{K}}^*=\left( \frac{c}{\pi R_{\rm{g}}} \right) \left[ \left(\frac{2}{r}\right)^{3/2} + a  \right]^{-1} ,
	\label{eq2}
\end{equation}

where $r=R_{\rm{g}}/R_{\rm{in}}$, $a$ is black hole spin, $M$ is the black hole mass, and $f_{\rm{K}}^*$ is the Keplerian frequency. The characteristic frequency $\nu_0$ is inversely proportional to the intrinsic soft lag amplitude, and further proportional to $1/R_{\rm{in}}$ under the truncated disk geometry \citep{de2021inner}. In order to testify whether the geometry suggested by the evolution of $\nu_0$ is consistent with the prediction of the L-T precession model, we multiply $1/r$ by a constant $A$ and use the deformed formula to fit the $\nu_0$-QPO frequency relation. The fitting result is given in Figure~\ref{LT} ($A=132\pm24$), assuming a black hole mass of ten solar masses and a spin of 0.998. $\nu_0$ and 1/ $R_{\rm{in}}$ show a high degree of consistency in terms of type-C QPO frequency. This correlation provides strong evidence of the L-T precession under the truncated disk geometry. It is worth noting that, the rigid body precession model is not used here because it involves many variable parameters. What we concern more is whether the $R_{\rm{in}}$ obtained by the reverberation mapping and the $R_{\rm{in}}$ predicted by the L-T precession model have the same evolutionary trends. Therefore, we used the simplified test-particle precession model to avoid the interference from other parameters. We intend to make a qualitative conclusion rather than strict quantitative calculations. 
 
In the SIMS, $\nu_0$ shows a dramatic increase in Figure~\ref{v0-relation}, implying that the relative distance between the disk and corona is significantly smaller than is the case in the HIMS. Previous studies have suggested that when the disk reaches the ISCO, the collapses of the inner flow could trigger a similar collapse of the radio emission \citep{done2007modelling}. The inner disk probably has reached the ISCO in the SIMS for this source, given that radio flares have been observed in the SIMS of this source \citep{chauhan2019h}. 

It is interesting to mention that a recent work, performed on the systematic study of reverberation lags of multiple BHXRBs, shows an opposite trend from ours on MAXI J1535-571 \citep{wang2022nicer}. The difference could be caused by the selections of energy bands, the criteria of state classification, or the method for measuring characteristic frequencies. In addition, the strong absorption of MAXI J1535-571 makes it more complicated to measure its soft lags and characteristic frequencies. It should be emphasized that the rebin factor can significantly change the directly measured values of $\nu_0$. Thus, we use a logarithmic function to measure the characteristic frequencies $\nu_0$. Our results also suggest that the soft lags evolution of MAXI J1535-571 seems to be different compared to MAXI J1820+070. \citet{de2021inner} also found that $\nu_0$ increases during the hard state but decreases during the transition to the soft state in MAXI J1820+070. They interpreted this as the emission from a ballistic jet becomes significant so that a larger area of the disk may be irradiated. MAXI J1535-571 also showed significant jet activity in the SIMS \citep{russell2019disk,russell2020rapid,vincentelli2021fast}, but $\nu_0$ increased compared to the HIMS. The different evolution trends of $\nu_0$ in the SIMS may indicate that MAXI J1535-571 has a different disk-corona geometry than MAXI J1820+070. In particular, the jet base of MAXI J1535-571 might remain close to the black hole during the SIMS, causing the hard photons to mainly irradiate the inner part of the disk.

Our spectral fitting results suggest that during the outburst, the inner disk moves inwards to the BH, from 38 $R_{\rm{ISCO}}$ to $<$ 2 $R_{\rm{ISCO}}$, which corresponds to the increase of $\nu_0$. In order to compare the relation between $R_{\rm{in}}$ and 1/$\nu_0$, we use a linear function to fit it (see Fig ~\ref{Rin_v0}). The figure shows that when $R_{\rm{in}}$ reaches the ISCO, 1/$\nu_0$ decreases significantly. These suggest that the accretion disk is truncated during the HIMS while reaches the ISCO during the SIMS.

It is worth noting that despite the high degeneracy between the corona height $h$ and the inner disk radius $R_{\rm{in}}$ (see Figure~\ref{fig12}), the lamppost model $relxilllpCp$ also prefers the disk to be truncated (see Table~\ref{tableA3}). A shrinking corona can also bring the changes in the reverberation mapping lags. However, it's hard to tell whether the corona is shrinking independently from $R_{\rm{in}}$ because of the parameter degeneracy. On the contrary, $relxillCp$ model doesn't show obvious degeneracy. However, if considering that the type-C QPOs are produced by the L-T precession, a truncated disk geometry is a more reasonable scenario, since a lamppost geometry can not give a precession corona. In conclusion, we prefer the interpretation with the lower height of the corona and the larger truncation of the inner disk. Of course, we cannot completely rule out the interpretations of other models.

\begin{acknowledgments}
This work has made use of the data from the \textit{Insight}-HXMT mission, a project funded by China National Space Administration (CNSA) and the Chinese Academy of Sciences (CAS), and data and/or software provided by the High Energy Astrophysics Science Archive Research Center (HEASARC), a service of the Astrophysics Science Division at NASA/GSFC. This work is supported by the National Key RD Program of China (2021YFA0718500) and the National Natural Science Foundation of China (NSFC)
under grants U1838201, U1838202, 11733009, 11673023, U1838111, U1838108, U1938102, U2038104, U1838110, U1838113, U1838115, U2031205, 12133007, and 12233002.
We thank Mariano Mendez and Yuexin Zhang for helpful discussions.
\end{acknowledgments}

%% To help institutions obtain information on the effectiveness of their 
%% telescopes the AAS Journals has created a group of keywords for telescope 
%% facilities.
%
%% Following the acknowledgments section, use the following syntax and the
%% \facility{} or \facilities{} macros to list the keywords of facilities used 
%% in the research for the paper.  Each keyword is check against the master 
%% list during copy editing.  Individual instruments can be provided in 
%% parentheses, after the keyword, but they are not verified.

%% Appendix material should be preceded with a single \appendix command.
%% There should be a \section command for each appendix. Mark appendix
%% subsections with the same markup you use in the main body of the paper.

%% Each Appendix (indicated with \section) will be lettered A, B, C, etc.
%% The equation counter will reset when it encounters the \appendix
%% command and will number appendix equations (A1), (A2), etc. The
%% Figure and Table counter will not reset.
\clearpage
\appendix
\setcounter{table}{0}
\setcounter{figure}{0}
\renewcommand{\thetable}{A\arabic{table}}
\renewcommand{\thefigure}{A\arabic{figure}}

\renewcommand\arraystretch{1}
\begin{table}[htbp]
	\begin{center}
	\caption{\emph{NICER} observation of MAXI J1535--571}
	\label{tableA1}
	\begin{tabular}{cccccccccc}
		\hline
		\hline
		ObsID  & MJD & obs time (ks) & $\nu_0$ & State & ObsID  & MJD & obs time (ks) & $\nu_0$ & State \\
		\hline
1050360104 &     58008.7  & 4.8  & $4.7\pm 0.3 $  &HIMS   &1050360118 &     58022.6  & 3.6  & $-$  &SIMS           \\
1050360105 &     58009.5  & 9.7  & $4.4\pm 0.7 $  &HIMS   &1050360119 &     58023.5  & 3.6  & $-$  &SIMS           \\
1050360106 &     58010.2  & 5.6  & $3.7\pm 0.7 $  &HIMS   &1130360101 &     58024.8  & 2.2  & $-$  &SIMS           \\
1050360107 &     58011.9  & 1.5  & $4.1\pm 0.5 $  &HIMS   &1130360102 &     58025.8  & 1.4  & $-$  &SIMS           \\
1050360108 &     58012.4  & 3.3  & $4.8\pm 0.3 $  &HIMS   &1130360103 &     58026.8  & 3.0  & $17.9\pm 1.4 $  &HIMS           \\
1050360109 &     58013.6  & 4.0  & $6.7\pm 1.7 $  &HIMS   &1130360104 &     58027.8  & 1.5  & $11.1\pm 4.7 $  &HIMS           \\
1050360110 &     58014.4  & 3.0  & $7.9\pm 1.9 $  &HIMS   &1130360105 &     58028.8  & 5.1  & $16.1\pm 1.2 $  &HIMS           \\
1050360111 &     58015.5  & 1.9  & $15.5\pm 7.1 $  &HIMS  &1130360106 &     58029.8  & 3.7  & $11.2\pm 2.4 $  &HIMS            \\
1050360112 &     58016.6  & 2.7  & $-$  &SIMS     &1130360107 &     58030.8  & 3.9  & $14.2\pm 1.1 $  &HIMS         \\
1050360113 &     58017.4  & 5.6  & $-$  &SIMS     &1130360108 &     58031.6  & 3.4  & $12.1\pm 1.5 $  &HIMS         \\
1050360114 &     58018.7  & 2.1  & $-$  &SIMS     &1130360110 &     58033.4  & 6.4  & $16.4\pm 1.6 $  &HIMS         \\
1050360115 &     58019.5  & 9.2  & $-$  &SIMS     &1130360112 &     58035.6  & 11.8  & $11.8\pm 4.2 $  &HIMS         \\
1050360116 &     58020.3  & 6.3  & $-$  &SIMS     &1130360113 &     58036.5  & 3.1  & $12.6\pm 2.9 $  &HIMS         \\
1050360120 &     58020.8  & 3.6  & $-$  &SIMS     &1130360114 &     58037.4  & 1.3  & $11.6\pm 4.2 $  &HIMS         \\
1050360117 &     58021.5  & 4.7  & $-$  &SIMS     & &   &  &   &        \\

		\hline
	\end{tabular}
	\end{center}
\end{table}

\renewcommand\arraystretch{1}
\begin{table}[htbp]
	\centering
	\caption{\emph{Insight}-HXMT observation of MAXI J1535-571}
	\label{tableA2}
	\begin{tabular}{cccccccccc} 
		\hline
		\hline
		$^a\rm{ObsID}$    & Date & MJD & obs time (ks)  & State & $^a\rm{ObsID}$ & Date & MJD & obs time (ks) & State \\
		\hline
106 &   2017-09-06   &  58002.5  &   11   &LHS      & 905 &                         &  58017.9  &   11   &SIMS    \\
107 &                         &  58002.6  &   11   &LHS   & 906 &   2017-09-22   &  58018.0  &   12   &SIMS       \\
108 &                         &  58002.7  &   11   &LHS   & 907 &                         &  58018.2  &   12   &SIMS       \\
144 &   2017-09-12   &  58008.4  &   12   &HIMS    &908 &                         &  58018.3  &   12   &SIMS    \\
145 &                         &  58008.6  &   38   &HIMS  &909 &                         &  58018.4  &   32   &SIMS      \\
301 &   2017-09-15   &  58011.2  &   11   &HIMS   &910 &                         &  58018.8  &   10   &SIMS     \\
401 &   2017-09-16   &  58012.3  &   11   &HIMS   &911 &                         &  58018.9  &   11   &SIMS     \\
501 &   2017-09-17   &  58013.3  &   11   &HIMS   &912 &   2017-09-23   &  58019.1  &   12   &SIMS     \\
601 &   2017-09-18   &  58014.1  &   11   &HIMS   &913 &                         &  58019.2  &   12   &SIMS     \\
701 &   2017-09-20   &  58016.0  &   12   &SIMS   &914 &                         &  58019.4  &   12   &SIMS     \\
901 &   2017-09-21   &  58017.1  &   11   &SIMS   &915 &                         &  58019.5  &   12   &SIMS     \\
902 &                         &  58017.2  &   12   &SIMS  &916 &                         &  58019.6  &   11   &SIMS      \\
903 &                         &  58017.4  &   12   &SIMS  &917 &                         &  58019.8  &   11   &SIMS      \\
904 &                         &  58017.5  &   32   &SIMS  &918 &                         &  58019.9  &    9    &SIMS      \\

		\hline
	\end{tabular}
	\begin{tablenotes}
    \footnotesize
    \item{a} P011453500XXX
    \end{tablenotes}
\end{table}

\begin{figure*}[htbp]
	\centering\includegraphics[width=18cm]{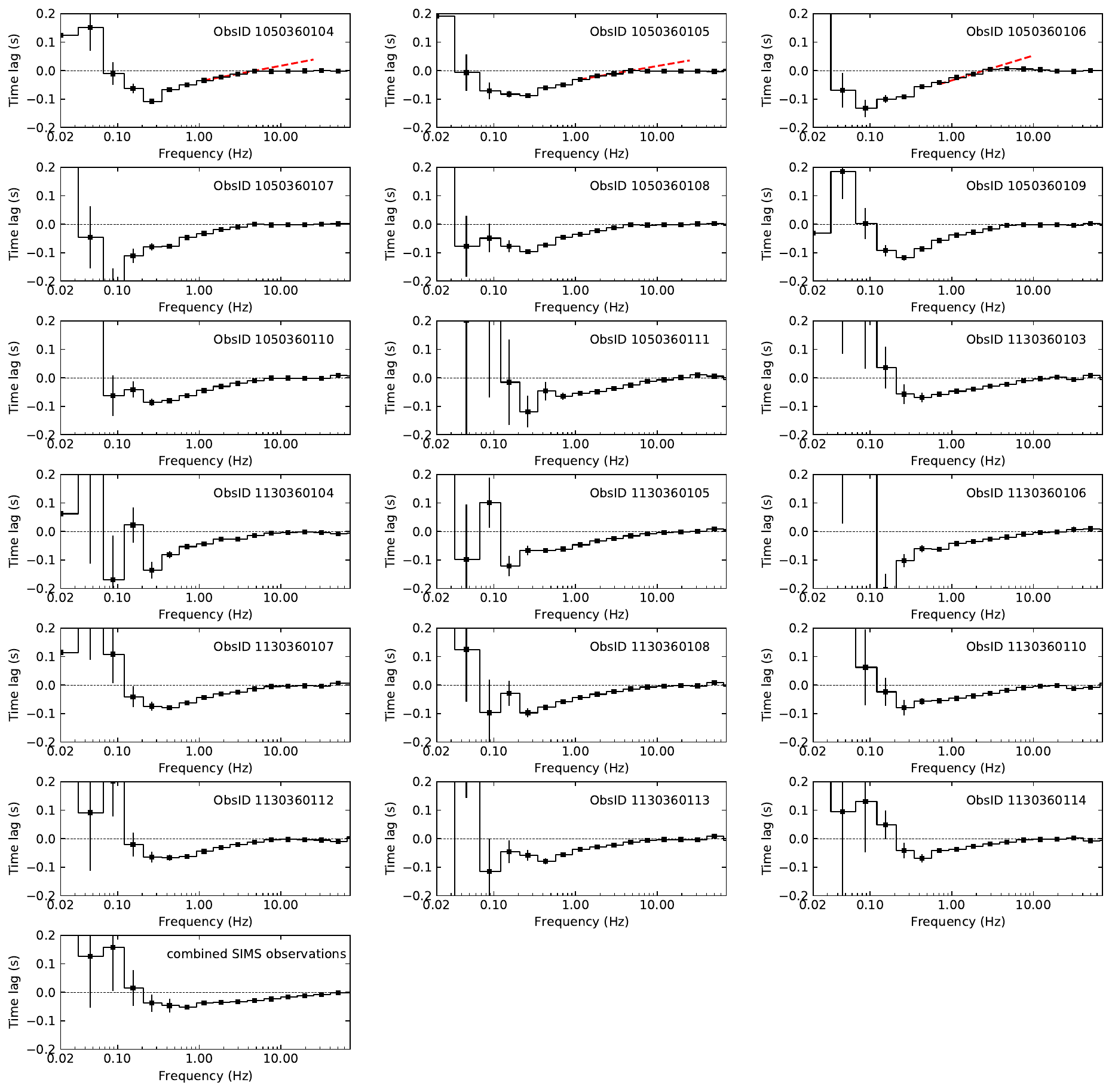}
    \caption{The 0.5–2.5 keV vs. 3–5 keV lag-frequency spectra of all selected NICER observations of MAXI J1535–571. The red dashed line represents the best logarithmic fit.}
    \label{ccf_all}
\end{figure*}

\begin{figure*}[htbp]
	\centering\includegraphics[width=18cm]{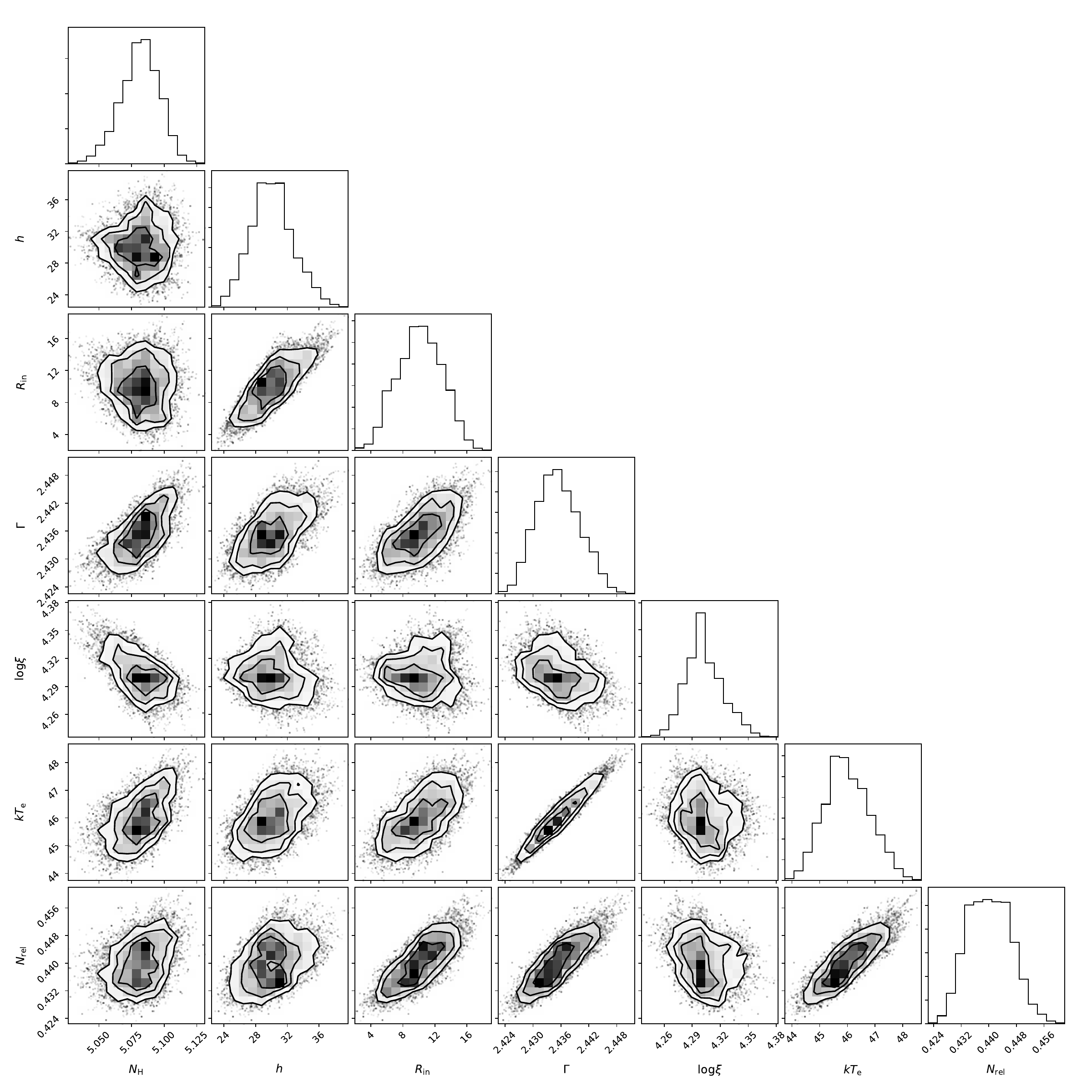}
    \caption{ Two dimensional projection of the posterior probability distributions derived from the MCMC analysis for the parameters in $relxilllpCp$. The contours plot 1-, 2- and 3-$\sigma$ confidence. This illustration corresponds to the spectral fitting of ObsID P011403500501. The figure is produced using the corner package \citep{corner}.}
    \label{fig12}
\end{figure*}
\FloatBarrier
\renewcommand\arraystretch{1.2}
\begin{table*}
    \begin{center}
	\caption{Best fitting parameters for model3 : $constant*TBabs*(relxilllpCp+diskbb)$}.
	    \setlength{\tabcolsep}{1mm}{
	    \begin{tabular}{cccccccccccc}
		    \hline
		    $^a\rm{ObsID}$ & $N_{\rm{H}}$ & $T_{\rm{in}}$ & $N_{\rm{disk}}$ & $h$  & $R_{\rm{in}}$ & $\Gamma$ & log$\xi$ & $kT_{\rm{e}}$  & $N_{\rm{rel}}$ & $\chi^{2}_{\rm{red}}(d.o.f)$\\
		     & ($10^{22}{\rm{cm}}^{-2}$) & (keV) &($10^3)$ & (Rg) & (ISCO) &  &  & (keV) &   \\
		    \hline
            106 & $4.40_{-0.8}^{+0.06} $ & $-$ & $-$ &$32_{-15}^{+22} $ & $20_{-12}^{+18} $ & $1.77_{-0.01}^{+0.01} $ & $3.51_{-0.02}^{+0.02} $ & $29.2_{-0.6}^{+1.1} $  & $0.08_{-0.01}^{+0.01} $ & 1.11(1207) \\
            145 & $4.62_{-0.05}^{+0.03} $ & $-$ & $-$ &$38_{-9}^{+17} $ & $25_{-14}^{+21} $ & $2.33_{-0.01}^{+0.01} $ & $4.02_{-0.05}^{+0.04} $ & $37.5_{-1.2}^{+2.1} $  & $0.27_{-0.03}^{+0.02} $ &1.12(1175) \\
            301 & $4.72_{-0.05}^{+0.03} $ & $-$ & $-$ &$25_{-11}^{+10} $ & $32_{-19}^{+10} $ & $2.27_{-0.01}^{+0.01} $ & $3.85_{-0.03}^{+0.04} $ & $25.1_{-1.5}^{+1.2} $  & $0.26_{-0.01}^{+0.01} $ &1.04(1205)\\
            401 & $4.78_{-0.04}^{+0.07} $ & $-$ & $-$ &$43_{-13}^{+21} $ & $23_{-14}^{+8} $ & $2.32_{-0.01}^{+0.01} $ & $4.12_{-0.05}^{+0.03} $ & $32.6_{-1.1}^{+0.9} $  & $0.34_{-0.01}^{+0.01} $ &1.12(1207)\\
            501 & $5.07_{-0.02}^{+0.04} $ & $-$ & $-$ &$29_{-5}^{+7} $ & $10.3_{-5.1}^{+6.3} $ & $2.43_{-0.01}^{+0.01} $ & $4.30_{-0.04}^{+0.05} $ & $45.8_{-1.5}^{+2.1} $  & $0.44_{-0.01}^{+0.01} $ &1.18(1173)\\
            601 & $5.01_{-0.06}^{+0.12} $ & $-$ & $-$ &$35_{-11}^{+5} $ & $16.4_{-8.2}^{+3.7} $ & $2.42_{-0.01}^{+0.01} $ & $4.19_{-0.03}^{+0.02} $ & $38.6_{-2.5}^{+1.8} $  & $0.43_{-0.02}^{+0.01} $ &1.11(1151)\\
            901 & $4.90_{-0.05}^{+0.09} $ & $1.20_{-0.01}^{+0.01}$ & $2.29_{-0.02}^{+0.04}$ &$17.2_{-4.2}^{+3.8} $ & $5.2_{-3.6}^{+5.1} $ & $2.56_{-0.01}^{+0.01} $ & $4.70_{-0.11}^{+0.13} $ & $233_{-15}^{+24} $  & $0.45_{-0.01}^{+0.02} $ &1.12(1205)\\
            906 & $4.55_{-0.06}^{+0.05} $ & $1.16_{-0.01}^{+0.01} $ & $4.29_{-0.06}^{+0.05} $ &$10.4_{-2.4}^{+5.3} $ & $11.6_{-8.1}^{+4.6} $ & $2.79_{-0.01}^{+0.01} $ & $3.11_{-0.12}^{+0.17} $ & $400_{-21}^{+0} $  & $0.52_{-0.01}^{+0.02} $ &1.08(1205)\\
            912 & $4.45_{-0.08}^{+0.09} $ & $1.16_{-0.01}^{+0.02}$ & $4.17_{-0.05}^{+0.07}$ &$15.3_{-3.5}^{+6.3} $ & $6.6_{-3.2}^{+4.7} $ & $2.73_{-0.01}^{+0.01} $ & $3.09_{-0.12}^{+0.15} $ & $400_{-17}^{+0} $  & $0.53_{-0.02}^{+0.01} $ &1.05(1205)\\
		    \hline
	    \end{tabular}
        }
        \label{tableA3}
	\begin{tablenotes}
        \footnotesize
        \item{a} P011453500XXX
        \end{tablenotes}
    \end{center}
\end{table*}

%% For this sample we use BibTeX plus aasjournals.bst to generate the
%% the bibliography. The sample631.bib file was populated from ADS. To
%% get the citations to show in the compiled file do the following:
%%
%% pdflatex sample631.tex
%% bibtext sample631
%% pdflatex sample631.tex
%% pdflatex sample631.tex

\bibliography{sample631}{}
\bibliographystyle{aasjournal}

%% This command is needed to show the entire author+affiliation list when
%% the collaboration and author truncation commands are used.  It has to
%% go at the end of the manuscript.
%\allauthors

%% Include this line if you are using the \added, \replaced, \deleted
%% commands to see a summary list of all changes at the end of the article.
%\listofchanges

\end{document}